\documentclass[twocolumn,aps,pra,showpacs,raggedbottom,nobalancelastpage,amssymb,superscriptaddress]{revtex4-1}

\usepackage{graphicx}
\usepackage{amsmath}
\usepackage{amssymb}
\usepackage{bm}
\usepackage[usenames]{color}
\usepackage{amsfonts}
\usepackage{appendix}
\usepackage{dsfont}

\usepackage[colorlinks=true , citecolor=blue,urlcolor=blue]{hyperref}

\definecolor{Nathanblue}{rgb}{0.96,0.24,0.00}

\definecolor{Nathanred}{rgb}{0.06,0.24,0.90}

\def\be{\begin{equation}}
\def\ee{\end{equation}}

\begin{document}
\title{Synthetic Dimensions for Cold Atoms from Shaking a Harmonic Trap}
\author{Hannah M. Price}
\email[]{hannah.price@unitn.it}
\affiliation{INO-CNR BEC Center and Dipartimento di Fisica, Universit\`{a} di Trento, I-38123 Povo, Italy}
\author{Tomoki Ozawa}
\affiliation{INO-CNR BEC Center and Dipartimento di Fisica, Universit\`{a} di Trento, I-38123 Povo, Italy}
\author{Nathan Goldman}
\email[]{ngoldman@ulb.ac.be}
\affiliation{CENOLI, Facult{\'e} des Sciences, Universit{\'e} Libre de Bruxelles (U.L.B.), B-1050 Brussels, Belgium}
\begin{abstract}
We introduce a simple scheme to implement synthetic dimensions in ultracold atomic gases, which only requires two basic and ubiquitous ingredients: the harmonic trap, which confines the atoms, combined with a periodic shaking. In our approach, standard harmonic oscillator eigenstates are reinterpreted as lattice sites along a synthetic dimension, while the coupling between these lattice sites is controlled by the applied time-modulation. The phase of this modulation enters as a complex hopping phase, leading straightforwardly to an artificial magnetic field upon adding a second dimension. We show that this artificial gauge field has important consequences, such as the counterintuitive reduction of average energy under resonant driving, or the realisation of quantum Hall physics. Our approach offers significant advantages over previous implementations of synthetic dimensions, providing an intriguing route towards higher-dimensional topological physics and strongly-correlated states.
\end{abstract}

\date{\today}
\maketitle

\section{Introduction}

Charged quantum particles in magnetic fields exhibit important phenomena, not least of which is the famous quantum Hall (QH) effect~\cite{QH}, in which the quantised Hall conductance is related to robust chiral edge states and non-zero topological band invariants~\cite{TKNN,RMP_TI, RMP_TI2}. Although first studied in solid-state materials, there has been great progress in exploring QH physics also with ultracold atoms~\cite{ReviewNG,jotzu2014,Aidelsburger:2015, Mancini:2015,Stuhl:2015,Wu2015}, photons~\cite{TopPhoton, Rechtsman, Hafezi}, classical circuits~\cite{Simon, Albert} and mechanical systems~\cite{HuberComment,Huber, Salerno:2015}, where the magnetic field effects must be engineered artificially~\cite{Dalibard2011,Goldman:2014bv, Hafezi_review}. 

An exciting development in this direction has been the introduction of the ``synthetic dimension" concept into ultracold atomic gases~\cite{Boada2012, Celi:2014, Mancini:2015,Stuhl:2015,CooperRey,Gadway,Gadway2,Livi:2016, Gadway3} and photonics~\cite{peano:2015, Luo:2015,4Dphotons:2015, Fan}. In this approach, discrete internal degrees of freedom are reinterpreted as labelling lattice sites along an additional \emph{synthetic} dimension; by externally-coupling together these degrees of freedom, particles can be understood to ``hop" between lattice sites and so move along this extra dimension. When combined with other (real) spatial dimensions, artificial magnetic fields can be imprinted by spatially varying the external coupling~\cite{Celi:2014}.

In ultracold gases, a synthetic dimension can be implemented by using different internal atomic states as the lattice sites, with an inter-site coupling and artificial magnetic flux controlled by external Raman lasers~\cite{Boada2012, Celi:2014}. Experimentally, this was demonstrated for atoms in a 1D optical lattice with three internal states, effectively simulating a three-leg ladder pierced by a uniform magnetic field~\cite{Mancini:2015,Stuhl:2015}. However, this approach faces significant challenges:~it is technically demanding to couple more internal states and so the synthetic dimension is extremely short; this scheme strongly relies on the available atomic species and on the practical constraints imposed by the atom-light coupling;  tunneling matrix elements along the synthetic dimension are  highly non-uniform due to Clebsch-Gordan coefficients associated with atomic transitions; and interactions are typically independent of the internal state and so of the ``separation" in the synthetic dimension. This contrasts with real dimensions, where interactions decay with inter-particle distance. In such synthetic dimensions, the state-independent (``zero-dimensional") interactions preclude the realization of fractional QH liquids~\cite{Lacki:2016}.  

\subsection{Scope of this work}

\begin{figure}[!]
\resizebox{0.48\textwidth}{!}{\includegraphics*{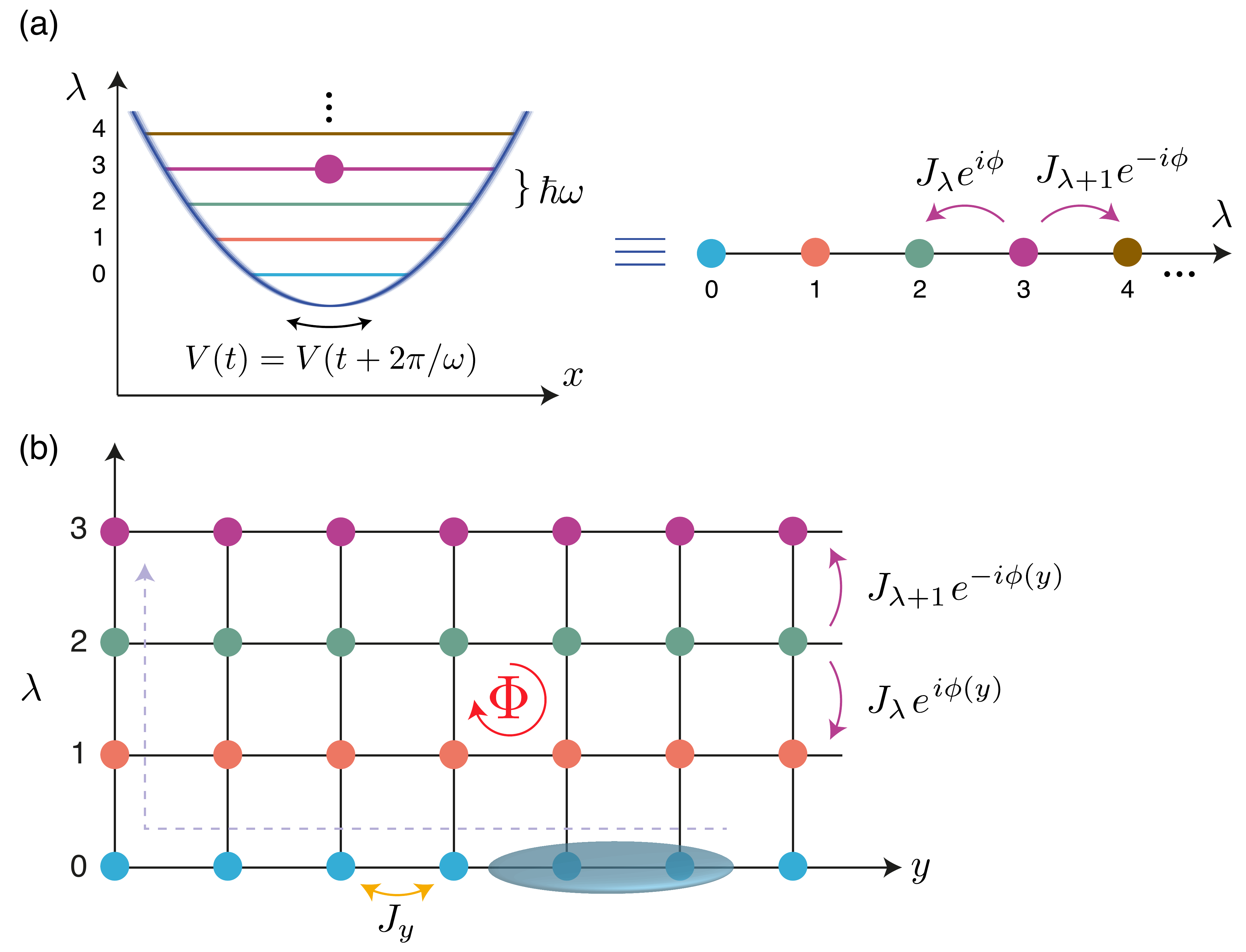}}  
\caption{(a) A particle in a periodically-shaken harmonic trap [Eq.~\eqref{Hzero}-\eqref{modulation_basis}] can be understood as moving along a synthetic dimension indexed by the HO number $\lambda$ [Eq.~\ref{effective_ham}]. The modulation phase $\phi$ enters as a complex hopping phase-factor, and so can be used to create artificial magnetic fields when more dimensions are added as in (b). In this configuration,  an atomic cloud initially prepared in the lowest-lying harmonic oscillator states ($\lambda\!\approx\!0$)  undergoes a unidirectional motion along the ``real" dimension $y$, until it reaches the edge at $y\!=\!0$; after reaching this point, the motion along $y$ stops, and the average energy increases as the atoms are driven to higher harmonic oscillator states. This motion is analogous to the unidirectional propagation of chiral edge states in 2D quantum Hall systems.}
 \label{fig:1} 
\end{figure}

In this work, we introduce a simple and versatile scheme for implementing synthetic dimensions, which only requires two basic ingredients: the harmonic potential, which is anyway present in experiments to confine the atoms, combined with a periodic shaking of the trap~\cite{Lignier,Oberthaler,Struck}. Here, sites along the synthetic dimension are indexed by the harmonic-oscillator  (HO) eigenstates quantum number $\lambda$, while motion between sites is controlled by the time-modulation [Fig.~\ref{fig:1}~(a)]. The phase of this modulation can be used to straightforwardly imprint an artificial magnetic field when the system is extended by a second dimension, e.g.~by a 1D optical lattice as shown in Fig.~\ref{fig:1}~(b). 

We demonstrate that this artificial gauge field has important consequences, such as an unusual reduction in the average energy despite resonant driving (as illustrated in Section~\ref{section_reduction}),  or robust one-way transport. For example, consider a cloud of atoms that is suitably prepared in the lowest-lying harmonic oscillator states ($\lambda\!\approx\!0$) and trapped in some region of an optical lattice directed along a transverse ($y$) direction [Fig.~\ref{fig:1}~(b)]: due to the artificial magnetic field defined in the abstract $\lambda-y$ plane, the cloud will undergo a chiral motion along the ``real" 1D lattice, as indicated in Fig.~\ref{fig:1}~(b). Only once the atoms reach the ``corner" of the system at $y\!=\!0$, will the cloud stop its propagation along $y$ and will the average energy begin to increase as the atoms are driven to higher harmonic oscillator states. This is directly analogous to the motion of a wave packet propagating chirally around the edge of a 2D quantum Hall system (here defined in the abstract $\lambda-y$ plane).

Our scheme is very general and will be directly applicable in ultracold atoms~\cite{ColdAtoms}. Furthermore, it offers practical advantages over many current QH ultracold gas experiments~\cite{jotzu2014,Aidelsburger:2015, Mancini:2015,Stuhl:2015,Wu2015} as it does not require the use of addressable internal states nor 2D (super)lattice potentials. This scheme will be well suited to measuring quantized Hall conductivity in cold-atom setups with constriction-based channels~\cite{Krinner}, where a 1D optical lattice may be easily designed. Our proposed implementation also overcomes the aforementioned limitations:~many HO eigenstates can be potentially coupled, leading to a long synthetic dimension; this scheme is independent of the available atomic species;  tunneling matrix elements are approximately uniform in the large $\lambda$ limit; and the interactions decay with the  ``separation" along the synthetic dimension, raising the possibility of accessing genuine 2D correlated states (e.g.~fractional QH states) when combined with a real dimension. Our proposal also opens a way to practically realize higher-dimensional topological physics~\cite{IQHE4D, 4Datoms:2015, 4Dphotons:2015, lian:2016}, even up to six spatial dimensions~\cite{Ryu}. We note that the dimensional crossover associated with the progressive population of harmonic-trap modes was recently investigated in Ref.~\cite{Lang:2016}. \\

\subsection{Outline}

The structure of this paper is as follows: in Section~\ref{sec:syn}, we show how harmonic oscillator states can be coupled by shaking and re-interpreted as lattice sites along a synthetic dimension. After deriving this picture in Sections~\ref{sec:syn_model} and \ref{sect:effective}, we justify its validity in Section~\ref{section:validity} and illustrate how it provides an alternative way to understand the center-of-mass evolution of a wave packet in a shaken harmonic trap in Section~\ref{sec:COMshaken}. In Section~\ref{section:2D_conf}, we discuss how to add a second dimension, so as to realise either a two-leg ladder, as further studied in Section~\ref{sec:2leg}, or a 2D lattice, as explored in Section~\ref{sec:2D}. We demonstrate that an artificial magnetic field, imposed via the modulation phase, has important consequences, including the counterintuitive reduction of average energy under resonant driving, and the realisation of quantum Hall physics. In Section~\ref{sec:extra}, we discuss how to extend our approach to create an {\it extra} dimension by using superlattice structures, opening the way towards the realisation of higher-dimensional topological phases. In Section~\ref{sec:inter}, we present the form of inter-particle interactions along the synthetic dimension, and compare these to interactions in other schemes. Finally, in Section~\ref{sec:discussion}, we discuss the effects of experimental anharmonicity in the harmonic trap as well as feasible experimental parameters. 

\section{Synthetic Dimension from Shaking a Single Harmonic Trap} \label{sec:syn}

\subsection{The model} \label{sec:syn_model}

We consider particles of mass $M$ in a harmonic potential of frequency $\omega$, aligned along the $x$ direction. The single-particle eigenstates $\{ \vert \lambda \rangle \}$ are indexed by the quantum number $\lambda \!=\!0,1,2, \dots$, and form an infinite ladder of equispaced states~[Fig.~\ref{fig:1}], with energies  $\lambda\hbar \omega$ (up to a constant energy off-set). The corresponding (static) Hamiltonian is written as
\be
\hat H_0=\frac{\hat p^2_x}{2M} +\frac{1}{2} M \omega^2 \hat x^2 = \omega \sum_{\lambda=0}^{\infty} \lambda \vert \lambda \rangle \langle \lambda \vert,\label{Hzero}
\ee
where we have set $\hbar\!=\!1$. In the following, the different states indexed by $\lambda$ will be re-interpreted as lattice sites along a synthetic dimension [Fig.~\ref{fig:1}]. In this scheme, motion along $\lambda$ will not be independent with respect to the ``real" dimension $x$. However, as discussed in Sec.~\ref{sec:extra}, our approach may be used to generate a genuine \emph{extra} dimension by using suitable couplings along a 1D array of traps. By addressing HO states along each real spatial dimension $(x, y, z)$ separately, we could then extend the effective dimensionality even further, e.g.~reaching an effective 6D lattice using a 3D array of traps. 

To couple different $\lambda$ states and so to implement a ``hopping" along the synthetic dimension, we apply a time-modulated linear gradient: 
\begin{eqnarray}
\hat V(t) \!=\! \kappa \hat x \cos (\omega_D t + \phi),
\end{eqnarray}
which is (nearly) resonant with the harmonic trap frequency, $\omega_D\!=\!\omega \!-\!\Delta$, where $\Delta \ll \omega$ is a small detuning~\cite{Yukalov:1997}. Note that this driving term may equally describe that of a particle moving in a shaken harmonic trap, as viewed in a co-moving reference frame~\cite{Creffield:2016,Jean_notes}. From standard properties of HO states, this time-modulation reads 
\be
\hat V(t) = \frac{\kappa}{\sqrt{2 M \omega}} \cos (\omega_{\text{D}} t + \phi) \sum_{\lambda=1}^{\infty} \sqrt{ \lambda } \biggl ( \vert \lambda \rangle \langle \lambda -1 \vert  +  \text{h.c.} \biggr ) .  \label{modulation_basis}
\ee
For the sake of presentation, we will no longer specify the limits on the sums over $\lambda$.


\subsection{The effective Hamiltonian}\label{sect:effective}

We are interested in the dynamics associated with the total time-dependent Hamiltonian
\be
\hat H_{\text{tot}}(t)\!=\!\hat H_0+\hat V(t),\label{Ham_tot}
\ee
in the high-frequency regime ($\omega \rightarrow \infty$) specified below. To analyze this, we first perform a change-of-frame transformation, using the unitary operator
\be
\hat R (t)= \exp \left ( i t \omega_{\text{D}}\hat \lambda  \right ), \label{rotating_frame}
\ee
where $\hat \lambda = \sum_{\lambda} \lambda \vert \lambda \rangle \langle \lambda \vert$ is the analogue of the position operator in $\lambda$-space. 
In the corresponding rotating frame, the time-dependent Hamiltonian \eqref{Ham_tot} takes the form
\begin{align}
\hat{\mathcal{H}}_{\text{tot}}(t)=&\Delta \sum_{\lambda} \lambda \vert \lambda \rangle \langle \lambda \vert + \sum_{\lambda} J_{\lambda} \biggl ( \vert \lambda -1 \rangle \langle \lambda  \vert e^{i \phi}  + \text{h.c.} \biggr ) \notag\\
&+\sum_{\lambda} J_{\lambda} \biggl ( \vert \lambda -1 \rangle \langle \lambda \vert e^{-i \phi}e^{-2i t \omega_{\text{D}}}   + \text{h.c.} \biggr ),\label{transf_ham}
\end{align}
where we introduce the effective ``tunneling" amplitude
\be
J_{\lambda}=\kappa \sqrt{\frac{\lambda}{8 M \omega}}.\label{J_lambda}
\ee
At this stage, one notices that the system is characterized by two significantly different energy scales: (i) the tunneling amplitudes $J_{\lambda}$, which set the bandwidth of the effective band structure (in some region of interest, $\lambda\!=\!0,\dots, L_{\lambda}$), and (ii) the energy associated with the driving frequency $\hbar \omega_{\text{D}}$; we note that the small detuning $\Delta\!\ll\! \omega_{\text{D}}$ could be of the order of the $J_{\lambda}$'s. In the high-frequency limit, which can now be defined through the rigorous condition
\be
\omega_{\text{D}}\!\gg\! J_{\lambda}=\kappa \sqrt{\frac{\lambda}{8 M \omega}}, \qquad \omega_{\text{D}}\!\simeq\!\omega\!\gg\! \Delta , \label{highlimit}
\ee
one can safely make a rotating-wave approximation (RWA), which consists of averaging the time-dependent Hamiltonian \eqref{transf_ham} over one period of the driving [i.e.~one neglects the fast-oscillating terms in the second line of Eq.~\eqref{transf_ham}]. This results in the effective Hamiltonian
\begin{align}
\hat{\mathcal{H}}_{\text{eff}}=&\Delta \sum_{\lambda} \lambda \vert \lambda \rangle \langle \lambda \vert + \sum_{\lambda} J_{\lambda} \biggl ( \vert \lambda -1 \rangle \langle \lambda  \vert e^{i \phi}  + \text{h.c.} \biggr ),\label{effective_ham}
\end{align}
which is directly analogous to the Hamiltonian of particles hopping in a 1D tight-binding lattice indexed by $\lambda$, where the detuning $\Delta$ provides an effective ``force" along the chain, and the modulation phase $\phi$ leads to  complex (Peierls) phase-factors~\cite{Hofstadter}; as shown below, this can be used to simulate artificial magnetic fields in more dimensions. In particular, we note that the phase $\phi$ can be used to invert the sign of the tunneling.

The limit of uniform hoppings $J_{\lambda}\!\approx\!J_{\lambda_0}$ is recovered at sufficiently large $\lambda_0$; however, as we now show, the RWA breaks down when $J_\lambda\!\sim \!\omega_D$ and so we must ensure that $\kappa$ is sufficiently small that the RWA is applicable to all significantly populated modes. 

\begin{figure*}[!]
\includegraphics[width=17cm]{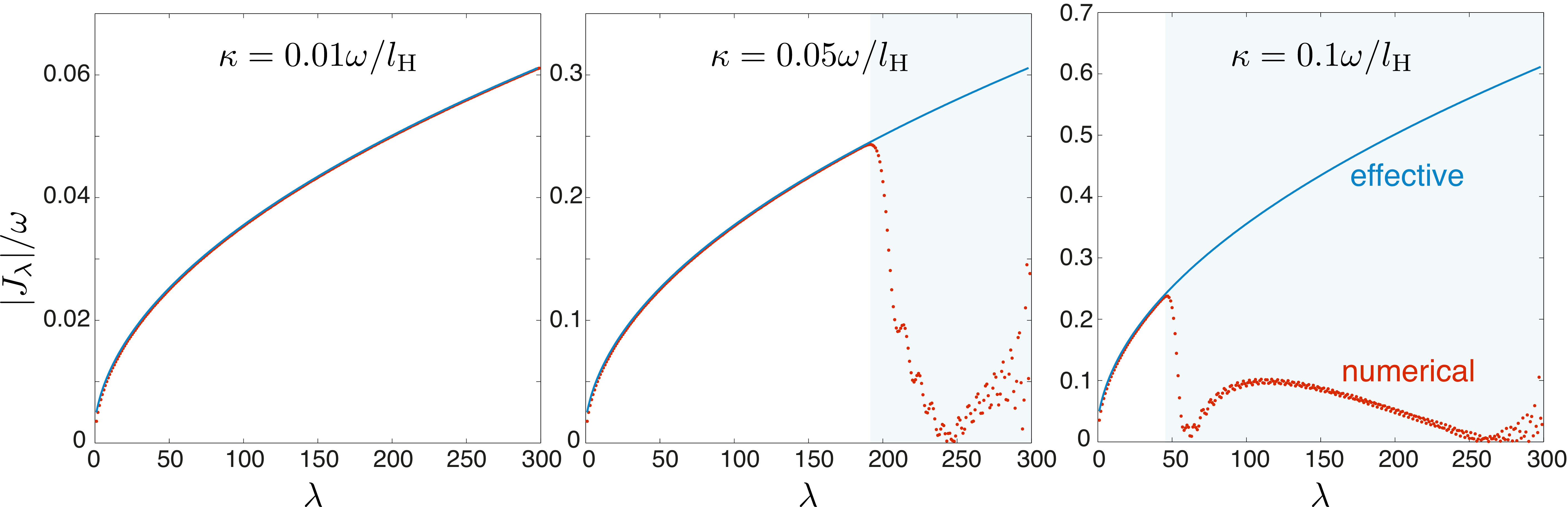}
\vspace{-0.cm} \caption{Comparison between the tunneling matrix elements $\vert J_{\lambda}\vert$ of the Floquet Hamiltonian in Eq.~\eqref{Floquet} (red dots), and the analytical expression \eqref{J_lambda} associated with the effective Hamiltonian in Eq.~\eqref{effective_ham} (blue line). The breakdown of the rotating-wave approximation \eqref{highlimit} is visible when $\vert J_{\lambda}\vert \gtrsim \omega/4$. The numerical data was obtained by splitting the time-evolution operator  $\hat U (T_{\text{D}};0)$ into 200 time steps, each of duration $\delta_t\!=\!T_{\text{D}}/200$, and within which the Hamiltonian was assumed to be constant; after evaluating $\hat U (T_{\text{D}};0)$, the Floquet Hamiltonian matrix is obtained through the expression $\hat{H}_{F}\!=\! i/T_{\text{D}} \log [ \hat U (T_{\text{D}};0)]$, which can be numerically evaluated. Here, we set the detuning $\Delta\!=\!0$.}\label{Fig_RWA}\end{figure*}

\subsection{Validity of the Effective Hamiltonian}\label{section:validity}

To demonstrate the validity of the effective Hamiltonian [Eq.~\eqref{effective_ham}], we introduce the stroboscopic time-evolution operator
\be
\hat U (NT_{\text{D}};0)=\left [ \hat U (T_{\text{D}};0) \right ]^N \approx  e^{- i NT_{\text{D}} \hat{\mathcal{H}}_{\text{eff}}} ,
\ee
where $\hat{U}(t_f; t_0)$ is the evolution operator from an initial time $t_0$ to a final time $t_f$, $N$ is some (arbitrarily large) integer, $T_{\text{D}}\!=\!2\pi/\omega_{\text{D}}$ is the period of the drive and $\hat{\mathcal{H}}_{\text{eff}}$ is the effective (RWA) Hamiltonian given in Eq.~\eqref{effective_ham}. Here we have neglected the effects of micro-motion~\cite{Goldman:2014, Goldman:2015}; these effects will be discussed below in Sec.~\ref{sec:micromotion}. Now introducing the exact ``Floquet" Hamiltonian $\hat{H}_{F}$ associated with the full time-dependent Hamiltonian~\eqref{Ham_tot}, the time-evolution operator is
\be
\hat U (T_{\text{D}};0)=e^{- i T_{\text{D}} \hat{H}_{F}}. \label{Floquet}
\ee
We now investigate the validity of the RWA [Eq.~\eqref{highlimit}], by numerically evaluating the matrix elements of $\hat{H}_{F}$ and comparing them to the analytical effective Hamiltonian $\hat{\mathcal{H}}_{\text{eff}}$ in Eq.~\eqref{effective_ham}. The results are shown in Fig.~\ref{Fig_RWA}, indicating the breakdown of the rotating-wave-approximation \eqref{highlimit} as soon as $\vert J_{\lambda}\vert \gtrsim \omega/4$. This figure also shows that the number of addressable $\lambda$ states (i.e.~the length of the synthetic lattice $L_{\lambda}$) is effectively set by the ratio $\kappa  l_{\text{H}}/\omega$ for an ideal harmonic potential, where we have introduced the HO length $l_{\text{H}} = 1/\sqrt{M \omega}$. When considering a reasonably large lattice size, $L_{\lambda}\approx 40$, Fig.~\ref{Fig_RWA} indicates that the RWA should remain valid as long as $\kappa \lesssim 0.1 \omega / l_{\text{H}}$. Note that, in practice, $L_{\lambda}$ will also be limited by anharmonicity in the trapping potential:~in experiments, this is typically set by the waist of Gaussian laser beams, as discussed further in Section~\ref{sec:discussion}.

\subsection{Center-of-mass evolution of a wave packet in a shaken harmonic trap}\label{sec:COMshaken}

To further verify the effective Hamiltonian [Eq.~\eqref{effective_ham}], we study the time-evolution in this shaken harmonic trap of a Gaussian wave packet $|\psi \rangle = \sum_{\lambda=0}^{\infty} \psi_{\text{gauss}} (\lambda) |\lambda \rangle \langle \lambda |$, where initially
\begin{equation}
\psi_{\text{gauss}} (\lambda)\!\propto\!e^{ -  (\lambda-\lambda_0)^2 / 2 \sigma^2 } , \label{eq:gauss}
\end{equation}
corresponding to a wave packet prepared around $\lambda_0$ with a width $\sigma$. Note that if $\lambda_0$ is insufficiently large or the spread $\sigma$ insufficiently small, then the wave packet can be strongly affected by the ``hard wall" at $\lambda\!=\!0$. In this case, the initial wave packet is not well-described by a Gaussian and $\lambda_0$ will not be equal to the mean of the populated-state distribution. Experimentally, cold atoms will typically be thermally distributed around $\lambda_0\!\approx\!0$; however, a well-defined Gaussian wave packet around $\lambda_0\!\gg\!0$ may be prepared using a well-designed pulse~\cite{JPBrantut}. 

\begin{figure}[b!]
\resizebox{0.45\textwidth}{!}{\includegraphics*{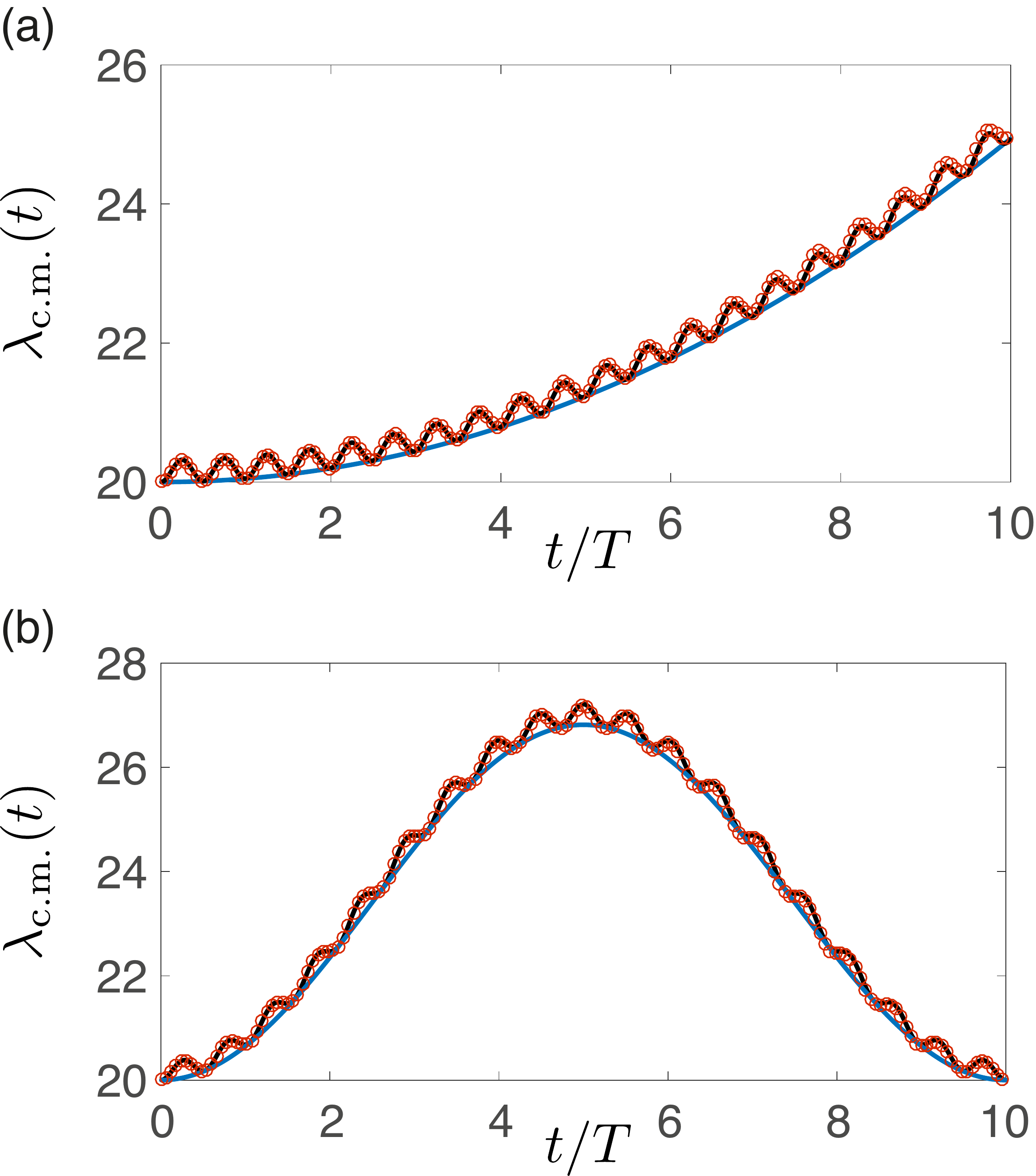}}
\caption{Numerical COM evolution $\lambda_{\text{c.m.}}(t)$ of a Gaussian wave packet for the 1D lattice model [Eq.~\ref{effective_ham}] (blue line), for the full Hamiltonian $\hat{H}(t) \!=\!  \hat{H}_0\!+\! \hat V(t)$ (red circles) and for the analytical classical COM evolution [Eq.~\ref{eq:analytic}] (black line), under a ``force" aligned along the synthetic dimension, as generated by the modulation detuning $\Delta$. (a) When $\Delta=0$, there is a run-away increase in $\lambda_{\text{c.m.}}(t)$ and hence of the average energy $E_{\text{av}}(t)\!=\!\omega{\lambda}_{\text{c.m.}} (t)$. (b) For small detuning, here $\Delta \!=\! 0.1 \omega$, there are large Bloch oscillations. In both panels, $\kappa \!=\!0.1 \omega / l_{\text{H}}$, $\phi\!=\!0$ and the wave packet is initialized with $\lambda_0\!=\!20$ and $\sigma\!=\!5$. Time is expressed in HO periods $T\!=\!2\pi/\omega$.} \label{fig:classical}
\end{figure}

In Fig.~\ref{fig:classical}, we numerically compare the center-of-mass (COM) evolution of the wave packet:
\begin{eqnarray}
{\lambda}_{\text{c.m.}} (t) \!=\! \langle \psi (t) | \hat{\lambda} |\psi (t) \rangle, 
\end{eqnarray}
under the full Hamiltonian $\hat{H}\!=\!\hat{H}_0\!+\!\hat{V}(t)$ (red circles) with the effective 1D hopping model [Eq.~\eqref{effective_ham}] (blue line). Here, we clearly identify two regimes; in the first regime [Fig.~\ref{fig:classical}(a)], the modulation is resonant ($\Delta = 0$) and there is a run-away increase in the average energy, $E_{\text{av}}(t)\!=\!\omega{\lambda}_{\text{c.m.}} (t)$. This is because $J_{\lambda+1} \!>\! J_{\lambda}$ in Eq.~\eqref{effective_ham} and so the COM increases over time. 

In the second regime [Fig.~\ref{fig:classical}(b)], the detuning is small, corresponding to a small external force. The wave packet then undergoes Bloch oscillations along $\lambda$, with a decreasing amplitude as $\Delta$ increases. This can be derived analytically from the 1D hopping model [Eq.~\eqref{effective_ham}] in the limit of large $\lambda_0$, in which hoppings along the synthetic dimension become effectively isotropic $J_\lambda\! \approx \!J_{\lambda_0}$. This model then has the usual energy dispersion of a tight-binding chain: $\varepsilon(q_\lambda) \!\approx \! 2 J_{\lambda_0} \cos ( q_\lambda)$ where $q_\lambda$ is the ``conjugate momentum" to $\lambda$. Under a force of $- \Delta$, a particle in this dispersion executes Bloch oscillations as described by:
\begin{eqnarray}
\lambda_{\text{c.m.}}(t)  = \lambda_0  + \frac{2 J_{\lambda_0}}{\Delta } \left(1   - \cos ( \Delta t) \right) ,\label{eq:BO}
\end{eqnarray}
where we impose the initial conditions that $ {\lambda_{\text{c.m.}}}(t\!=\!0)\! =\! \lambda_0$ and $q_\lambda(t\!=\!0)\!=\!0$. These Bloch oscillations correspond to the large-scale slow oscillations observed in Fig.~\ref{fig:classical}(b).  

However, while there is good large-scale agreement in Figs.~\ref{fig:classical}(a)\&(b), we also observe extra small oscillations in the full numerics. This is because we have chosen a large relative driving strength of $\kappa \!=\!0.1 \omega / l_{\text{H}}$, and so are not deep within the RWA limit [see Figure~\ref{Fig_RWA}]. In this case, we observe additional dynamical effects that go beyond the effective Hamiltonian [Eq.~\eqref{effective_ham}], but that can be captured instead by either a classical analysis, as we now derive in Sec.~\ref{sec:classical}, or by including micro-motion effects associated with the full Hamiltonian dynamics, as we present in Sec.~\ref{sec:micromotion}. 

\subsubsection{Classical Analysis} \label{sec:classical}

\begin{figure}[!]
\resizebox{0.47\textwidth}{!}{\includegraphics*{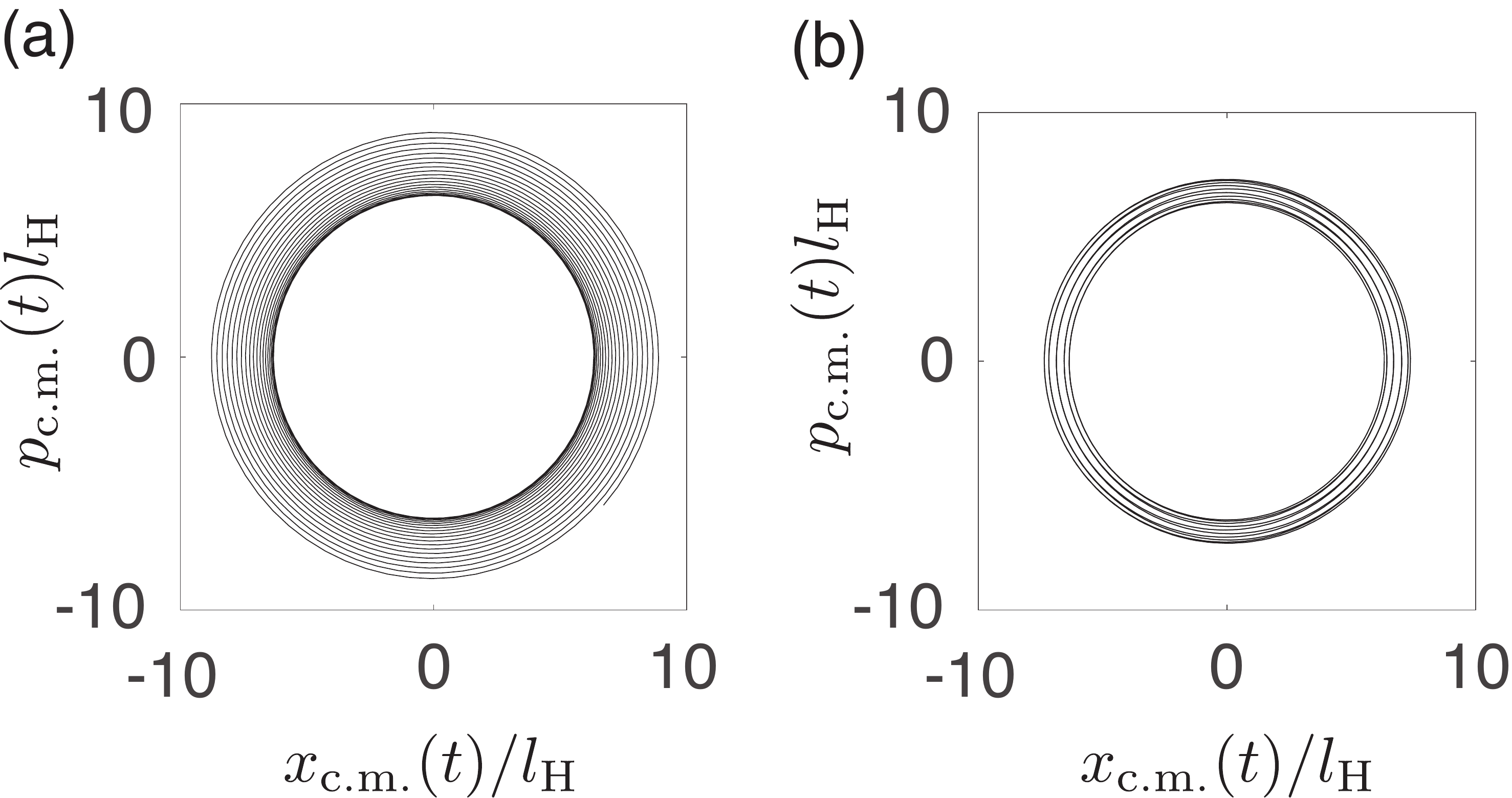}}
\caption{ Classical COM evolution of a Gaussian wavepacket up to $t\!=\!20 T$, as described by Eq.~\eqref{eq:analyticxp}, for two values of the modulation detuning $\Delta$. (a) Unstable trajectory in the case $\Delta\!=\!0$; this is to be compared with the diverging behavior of $\lambda_{\text{c.m.}}(t)$ in Fig.~\ref{fig:classical} (a). (b) Stable and periodic trajectory when $\Delta \!=\! 0.1 \omega$; this corresponds to large Bloch oscillations in $\lambda$-space [Fig.~\ref{fig:classical} (b)]. Parameters are the same as in Fig.~\ref{fig:classical}.} \label{fig:classical_traj}
\end{figure}

In this section, we show that a classical analysis can fully capture the COM behaviour of a Gaussian wave packet in a shaken harmonic trap. This approach provides analytical insight into the evolution and allows us to connect with the well-known classical physics of driven harmonic oscillators. Introducing $( x_{\text{c.m.}}, p_{\text{c.m.}})$ to denote the classical COM coordinates in phase space, the wave packet evolves according to:
\begin{eqnarray}
\dot{x}_{\text{c.m.}}& =& \frac{p_{\text{c.m.}}}{M}, \nonumber \\
\dot{p}_{\text{c.m.}}& =&  - M \omega^2 x_{\text{c.m.}} - \kappa \cos( (\omega- \Delta) t) ,\label{eq:classical_1}
\end{eqnarray}
where we have set $\phi\!=\!0$. These equations are, of course, the well-known equations of a driven harmonic oscillator, where the driving is detuned by an amount $\Delta$ from the harmonic oscillator frequency. Solving these analytically, we find:
\begin{eqnarray}
x_{\text{c.m.}}(t) &=& \frac{1}{M\Delta(\Delta- 2\omega)} \times \left[\kappa\cos((\Delta -\omega) t) \right. \nonumber\\
&& \left.+ (-\kappa + M x_1\Delta(\Delta-2\omega))\cos(\omega t))\right]\nonumber , \\
p_{\text{c.m.}}(t) &=& \frac{1}{\Delta(\Delta- 2\omega)} \left[ \kappa(\omega-\Delta)\sin((\Delta - \omega) t) \right. \nonumber \\
&& \left.+ (\kappa - M x_1 \Delta (\Delta-2\omega))\omega\sin(\omega t)\right],  \label{eq:analyticxp}
\end{eqnarray}
under the boundary conditions $p_{\text{c.m.}}(t\!=\!0)\!=\!0$ and $x_{\text{c.m.}}(t\!=\!0)\! =\! x_1$, where the initial position $x_1$ is specified below. These solutions reduce, as expected, to the usual HO evolution $x_{\text{c.m.}}(t) \!=\! x_1 \cos(\omega t) $ and $p_{\text{c.m.}}(t) \!=\!-  \omega x_1 \sin(\omega t)$ in the limit that the driving strength $\kappa $ vanishes. In Fig.~\ref{fig:classical_traj} (a)\&(b), we present the classical phase-space trajectories respectively for the two regimes shown in Fig.~\ref{fig:classical}. In the first regime of resonant driving, i.e.~$\Delta\!\rightarrow\!0$, we observe a run-away phase-space trajectory, as expected for a harmonic oscillator upon resonant driving, while in the second regime of Bloch oscillations, the center-of-mass evolves along a closed trajectory in phase space, corresponding to stable and periodic motion. 

To reinterpret this evolution in terms of motion along the synthetic dimension of HO states, we introduce the classical complex variable:
\begin{eqnarray}
\alpha_{\text{c.m.}} (t)  = \sqrt{\frac{M \omega} {2} } x_{\text{c.m.}} (t) + i \frac{p_{\text{c.m.}}(t)}{\sqrt{ 2 M \omega}} ,
\end{eqnarray}
from which it is straightforward to show that $ {\lambda_{\text{c.m.}}}(t)\! =\! |\alpha_{\text{c.m.}} (t)|^2$. For a Gaussian wave packet [Eq.~\ref{eq:gauss}], the initial COM position along the synthetic dimension is $ {\lambda_{\text{c.m.}}}(t\!=\!0)\! =\! \lambda_0$, which corresponds to the initial condition $x_1\! =\!\sqrt{2 \lambda_0 /  M \omega}$. Combining the above equations, we derive the full COM evolution as: 
\begin{eqnarray}
&& {\lambda_{\text{c.m.}}}(t)  =
\lambda_0 - \frac{\kappa^2 }{4 M \Delta \omega (\Delta - 2 \omega)} \cos( 2 (\Delta -  \omega)t)   \nonumber \\ &&
+ \frac{\kappa}{ (\Delta - 2 \omega)^2}\left( \sqrt{\frac{2 \lambda_0 \omega}{M}}
\left(   \frac{2\omega}{\Delta}-1 \right)   + \frac{\kappa (\Delta^2 - 2 \Delta \omega + 4 \omega^2)}{4 M \Delta^2 \omega} \right)      \nonumber \\ &&
  - \frac{\kappa}{( \Delta - 2 \omega) } \left( \frac{\kappa }{2 M \Delta  ( \Delta - 2 \omega)} 
- \sqrt{\frac{ \lambda_0}{ 2 M \omega } } \right)  \cos( (\Delta - 2 \omega)t) \nonumber \\ &&
+ \frac{\kappa }{ \Delta  }  \left( \frac{ \kappa}{2 M \Delta  ( \Delta - 2 \omega)  }  
- \sqrt{\frac{  \lambda_0}{ 2 M \omega} }    \right) \cos (\Delta t) . \label{eq:analytic}
\end{eqnarray}
As shown in Fig.~\ref{fig:classical}, this classical evolution (black line) is in excellent agreement with the quantum evolution of a wave packet under the full Hamiltonian (red circles). Furthermore, in the first regime of resonant driving, i.e.~$\Delta\!\rightarrow\!0$, the classical analytical solution becomes:
\begin{eqnarray}
 {\lambda_{\text{c.m.}}}(t) &=& \lambda_0+ \frac{\kappa^2}{8 M \omega} \left( t^2  +  \frac{t}{\omega}  \sin ( 2 \omega t) \right)
 \nonumber \\ &&  + \left( \frac{\kappa^2}{16 M \omega^3} + \frac{\kappa}{\omega } \sqrt{\frac{ \lambda_0 }{8 M \omega}} \right)
 \left( 1-  \cos(2 \omega t) \right) ,\qquad \label{eq:heating}
 \end{eqnarray}
which diverges with time as expected, and which contains additional oscillations at frequency $2 \omega$ as can also be clearly identified in the full numerics in Fig.~\ref{fig:classical}(a). As noted above, these extra oscillations are negligible in the RWA regime and so are not captured by the 1D effective hopping model. In the limit of small detuning, we recover the analytical Bloch oscillations [Eq.~\eqref{eq:BO}] from the classical evolution [Eq.~\eqref{eq:analytic}], along with additional oscillating terms that can be neglected in the RWA limit.

\subsubsection{The micro-motion and the initial kick} \label{sec:micromotion}

In this section, we briefly discuss how physics beyond the effective Hamiltonian [Eq.~\eqref{effective_ham}] can be captured in a quantum treatment by including the effects of micro-motion. The micro-motion associated with the full time-dependent Hamiltonian in Eq.~\eqref{transf_ham} is captured by the ``kick" operator~\cite{Goldman:2014, Goldman:2015}
\be
\hat{\mathcal K} (t)= \frac{i}{2  \omega_{\text{D}}}\sum_{\lambda} J_{\lambda} \biggl ( \vert \lambda -1 \rangle \langle \lambda \vert e^{-i (\phi+2 \omega_{\text{D}} t)}   - \text{h.c.} \biggr ).\label{kick}
\ee
In the rotating frame, the full time-evolution operator should then be approximated by the expression
\be
\hat U(t;t_0)=e^{-i \hat{\mathcal K} (t)} e^{- i (t-t_0) \hat{\mathcal{H}}_{\text{eff}}}e^{i \hat{\mathcal K} (t_0)}, \label{micro-partition}
\ee
with the effective Hamiltonian, $\hat{\mathcal{H}}_{\text{eff}}$, given in Eq.~\eqref{effective_ham}. We note that in the kick operator (Eq.~\eqref{kick}) we have only kept terms that are lowest order in $J_{\lambda}/\omega_D$ and $\Delta/\omega_D$ in the perturbative treatment. As the micro-motion does not depend on $\Delta$ at this order, we write $\omega_D\!=\!\omega$ in the remainder of this section [keeping in mind that $\Delta$ does appear in the effective Hamiltonian in Eqs.~\eqref{effective_ham} and \eqref{micro-partition}].

In order to estimate the effects of the micro-motion on wave-packet dynamics, let us introduce the lattice analogue of the momentum operator,
\be
\hat p_{\lambda}=\frac{1}{2i} \sum_{\lambda}  \vert \lambda -1 \rangle \langle \lambda \vert -\vert \lambda  \rangle \langle \lambda -1 \vert,
\ee
where we set $\phi\!=\!0$ for simplicity. Then, supposing that the dynamics are centered around some eigenstate of interest $\lambda\!\approx\! \lambda_0$, such that $J_{\lambda}\!\approx\! J_{\lambda_0}$, we note that the kick operator in Eq.~\eqref{kick} takes the simple form
\be
\hat{\mathcal K} (t)\approx - \frac{J_{\lambda_0}}{ \omega} \cos (2 \omega t) \hat p_{\lambda}+ \frac{J_{\lambda_0}}{2  \omega} \sin (2 \omega t) \hat T_{\lambda} ,\label{kick_simple}
\ee
where $\hat T_{\lambda}=\sum_\lambda \left (\vert \lambda -1 \rangle \langle \lambda \vert + \text{h.c.} \right )$ denotes the standard tunneling operator. The first term in Eq.~\eqref{kick_simple} indicates that the main effect associated with the micro-motion operator $\exp \left ( -i \hat{\mathcal K} (t) \right)$ corresponds to a translation of the wave-packet's center-of-mass, by an amount
\be
\lambda_{\text{c.m.}}  \rightarrow \lambda_{\text{c.m.}}  - \frac{J_{\lambda_0}}{ \omega} \cos (2 \omega t),\label{micro_trans}
\ee
at lowest order in $\kappa l_{\text{H}} / \omega $, while the second term in Eq.~\eqref{kick_simple} leads to an irrelevant deformation of the wave packet. This oscillating term in Eq.~\eqref{micro_trans} is found also in the classical analysis [Eq.~\eqref{eq:heating}] and is in good agreement with our numerical simulations of the full time dynamics. This underlines again how the micro-motion becomes irrelevant as one enters the high-frequency regime $J_{\lambda}\!\ll\!\omega_{\text{D}}$. 

Going further, we point out that the rotating-frame transformation in Eq.~\eqref{rotating_frame} also leads to a micro-motion when moving back to the original (laboratory) frame. Indeed, in that frame, the full time-evolution operator reads~\cite{Goldman:2014, Goldman:2015}
\be
\hat U(t;t_0)=\hat R^{\dagger} (t) e^{-i \hat{\mathcal K} (t)} e^{- i (t-t_0) \hat{\mathcal{H}}_{\text{eff}}}e^{i \hat{\mathcal K} (t_0)} \hat R(t_0) , \label{micro-partition2}
\ee
so that the micro-motion is now described by the product $\hat R^{\dagger} (t) \exp \left ( -i \hat{\mathcal K} (t) \right)$. Having studied the effects of the kick operator $\hat{\mathcal K} (t)$ above, we now focus on the subsequent effect associated with the operator $\hat R^{\dagger} (t)$. As can be shown from Eq.~\eqref{rotating_frame}, this additional micro-motion induces a shift $q_{\lambda}(t_f)\!=\! t_f \omega_{\text{D}}$ in the mean momentum of the wave packet, where $t_f$ indicates the time at which the system is probed. We note that this micro-motion in ``momentum" space is associated with the frequency $ \omega_{\text{D}}$ and does not affect the COM evolution of the wave packet along the synthetic dimension ${\lambda_{\text{c.m.}}}(t) $ that is studied above. 

Finally, as can be seen in Eq.~\eqref{micro-partition2}, time-dependent modulations also generate an initial kick~\cite{Goldman:2014}, as captured by the product of operators $\exp \left [ i \hat{\mathcal K} (t_0) \right]\hat R(t_0)$; see also Ref.~\cite{Goldman:2015}. As discussed above, these kicks correspond to sudden shifts in $\lambda$ and $q_{\lambda}$ space, which now depend on the initial time $t_0$ (instead of the final ``observation" time $t_f$). Such initial kicks can be suppressed or reduced through a proper choice of the initial time $t_0$ (i.e.~the initial phase of the modulation), or by slowly ramping up the time-modulation. In this work, we take $t_0\!=\!0$, such that the initial kick, $\exp \left [ i \hat{\mathcal K} (0) \right]$, only consists of a small center-of-mass shift $\lambda_{\text{c.m.}}(t_0) \rightarrow \lambda_{\text{c.m.}}(t_0) + (J_{\lambda_0}/\omega)$, where we have used that $\hat R(0)\!=\!\hat 1$. 
 
\section{Adding a second dimension} \label{section:2D_conf}

 By adding additional dimensions, we can use the harmonic oscillator states to engineer controllable and tuneable lattice models in 2D and higher. 

\begin{figure}[!]
\resizebox{0.49\textwidth}{!}{\includegraphics*{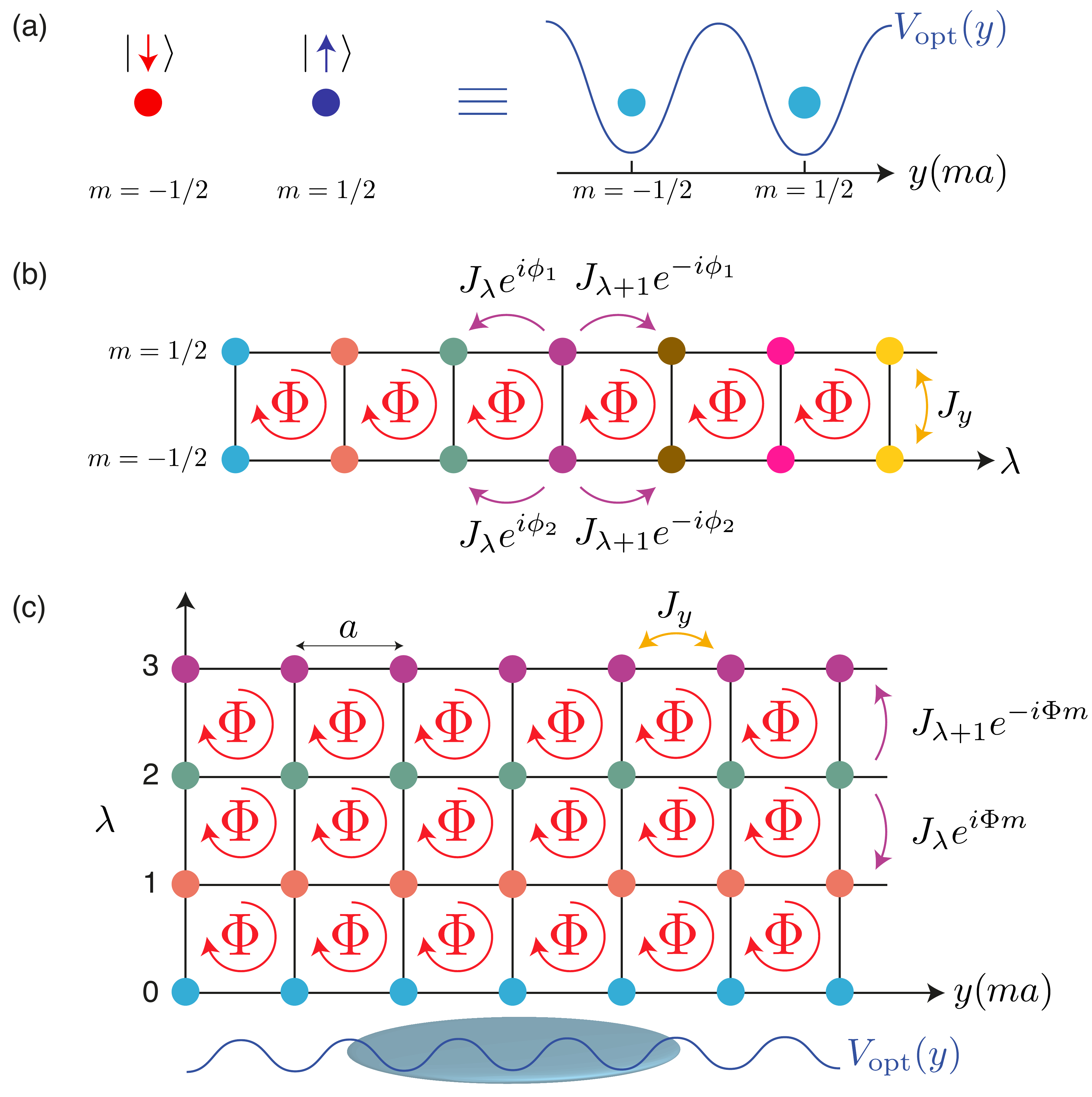}}  
\caption{(a) Two lattice sites along a second dimension, as given by (\emph{left}) two internal atomic states $\vert \uparrow, \downarrow \rangle$, or by (\emph{right})  a double well of spacing $a$ along $y$. Alternatively, the two lattice sites can correspond to two suitably-coupled micro-traps along $x$ [see Sec.~\ref{sec:extra}]. 
(b) Combined with the synthetic dimension $\lambda$, such a system can be viewed as a two-leg ladder pierced by a uniform magnetic flux $\Phi\!=\!\phi_2-\phi_1$, imposed by spin-dependent modulation phases $\phi_{1,2}$. 
(c) With more sites along the second dimension, e.g.~from an optical lattice along $y$, the system acts like a 2D (anisotropic) HH model. }
 \label{fig:2conf} 
\end{figure}

We illustrate the power of this approach here by showing how we can exploit the modulation phase $\phi$ to create an artificial magnetic field. To this end, we add a second dimension to our model as shown, for example, in Fig.~\ref{fig:2conf} and as discussed further below. Labelling sites along this second dimension by a site index, $m$, the total Hamiltonian becomes
\begin{align}
&\hat H_{\text{tot}}(t)= \omega \sum_{m, \lambda}  \lambda \vert \lambda , m \rangle \langle \lambda ,m \vert  - J_y \sum_{m,\lambda} \vert \lambda , m+1 \rangle \langle \lambda ,m \vert \notag\\
&+\frac{\kappa}{\sqrt{2 M \omega}} \sum_{m, \lambda} \cos (\omega_{\text{D}} t + \phi_m)  \sqrt{\lambda}   \vert \lambda ,m \rangle \langle \lambda -1 ,m \vert  +  \text{h.c.} , \label{H_tot_2D}
\end{align}
where $J_y$ denotes the tight-binding coupling along the second dimension. Following the same treatment as in Section~\ref{sect:effective}, the effective Hamiltonian can be written as
\begin{align}
\hat{\mathcal{H}}_{\text{eff}}&=\Delta \sum_{\lambda,m} \lambda \vert \lambda,m \rangle \langle \lambda,m \vert - J_y  \sum_{m,\lambda} \vert \lambda , m+1 \rangle \langle \lambda ,m \vert \notag\\
&+ \sum_{\lambda,m} J_{\lambda}  \vert \lambda -1,m \rangle \langle \lambda,m  \vert e^{i \phi_m}  + \text{h.c.}  \label{effective_ham_2D}  
\end{align}
which corresponds to the Harper-Hofstadter (HH) Hamiltonian, i.e.~a square lattice subjected to a uniform flux $ \Phi$, when the modulation phase $\phi_m$ is chosen as $\phi_m=  \Phi m + \theta$, where $\theta$ is an overall phase. We note that higher-order terms in the perturbative treatment lead to negligible higher-order hopping processes, such as next-nearest-neighbor (NNN) tunneling matrix elements of amplitude $J_{\lambda}^2/\omega$ [for NNN processes along $\lambda$], and ``crossed"-hopping terms of amplitude $J_y(J_{\lambda}/\omega)^2$  [for NNN processes involving $\lambda$ and $m$]; these negligible terms can be obtained from Eq.~\eqref{H_tot_2D}, via the method of Refs.~\cite{Goldman:2014, Goldman:2015}. 

In the following, we neglect the residual force due to the detuning ($\Delta\!=\!0$). However, we  note this detuning could be used to study Hall transport when both the magnetic flux $\Phi$ and force (artificial electric field) $\Delta$ are present. 

Interestingly, we note that had we considered a continuous dimension along $y$ instead of a lattice, i.e.~$\hat H_y\!=\!\hat p_y^2/2M$, then our system would be reminiscent of the quantum-Hall wires of Ref.~\cite{Kane2002}. In the present case, each wire is labelled by the mode index $\lambda$ and it is aligned along the continuous dimension $y$; the inter-wire couplings are given by $J_{\lambda}e^{i \phi (y)}$, leading to non-trivial topological bands~\cite{Kane2002}.

We now verify and explore the effective magnetic Hamiltonian in Eq.~\eqref{effective_ham_2D} for two relevant cases. Firstly, in Section~\ref{sec:2leg}, we consider the simplest experimental set-up where the second dimension consists of only two discrete lattice sites, labelled by $m\!=\!\{-1/2, +1/2\}$. In the following, we refer to this as a ``spin" degree of freedom, but physically, this could correspond to a deep double-well~\cite{Oberthaler} oriented along a second direction ($y$), perpendicular to the HO axis ($x$), or to two internal atomic states coupled together by external fields [Fig.~\ref{fig:2conf}(a)]. In terms of the effective Hamiltonian [Eq.~\eqref{effective_ham_2D}], a system with only two sites along the second dimension simulates a two-leg ladder pierced by a magnetic flux, as shown in Fig.~\ref{fig:2conf}(b). We note that we can also engineer such a configuration by coupling two micro-traps oriented along the HO axis ($x$); however, as the effective Hamiltonian is, generally, more involved, we delay further discussion of this to Sec.~\ref{sec:extra}. 

Secondly, in Section~\ref{sec:2D}, we explore the 2D QH regime by increasing the number of sites in the second dimension, such that generally $m\!\in\!\mathbb{Z}$. Physically, this could correspond to replacing the double-well along $y$ with a 1D optical lattice [Fig.~\ref{fig:2conf}(c)] or to coupling together more internal atomic states~\cite{Celi:2014}. Alternatively, as we again discuss further in Sec.~\ref{sec:extra}, we could also consider extending our system by a superlattice aligned along the HO axis ($x$), leading to a variant of the effective Hamiltonian in Eq.~\eqref{effective_ham_2D}.

\subsection{Chirality in a two-leg ladder}\label{sec:2leg}

In this section, we consider the two-leg ladder shown in Fig.~\ref{fig:2conf}(b), which corresponds to taking $m\!=\!\{-1/2, +1/2\}$ in Eq.~\ref{effective_ham_2D}. We choose the modulation phase $\phi_m$ to be $\phi_m= \Phi m + \theta$, such that there is a uniform flux of $\Phi$ per plaquette. Imposing the requisite hopping phases corresponds to varying the modulation phase $\phi$ between, e.g., the two wells or the two internal states in Fig.~\ref{fig:2conf}(a). Adjusting the overall phase $\theta$ adds a constant magnetic vector potential, which changes the gauge but not the magnetic flux.
A magnetic field generates chiral currents flowing along the two legs of a ladder~\cite{Hugel2014,Atala,Celi:2014,Piraud1,Piraud2}. This leads to remarkable phenomena when reinterpreted in terms of particles in a driven harmonic trap. For example, as we shall show, we can control the average energy of particles via the overall modulation phase $\theta$ (or equally, by the launching time of the drive~\cite{Goldman:2014}). 

\subsubsection{Energy dispersion of the isotropic two-leg ladder}

We begin by neglecting the hopping anisotropy (i.e.~setting $J_{\lambda}\!=\!J\!$) in Eq.~\eqref{effective_ham_2D} to derive the energy spectrum of the two-leg ladder in terms of the ``conjugate momentum" $q_\lambda$, as shown in Fig.~\ref{fig:2leg}(a). As the hopping amplitudes in our model increase with $\sqrt{\lambda}$ [Eq.~\eqref{J_lambda}], the isotropic case can be effectively recovered for regions at high $\lambda$, provided that the distribution of populated states is sufficiently well-localised within a given region. In particular, we approximate Eq.~\eqref{effective_ham_2D} as being perfectly isotropic within a region $\mathcal{R}$ consisting of $N$ sites along $\lambda$. Truncating the Hamiltonian to this region and Fourier-transforming, we find that $\hat{H}\! =\! \sum_{q_\lambda} \hat H_{q_\lambda}$, where
\begin{eqnarray}
\hat H_{q_\lambda} =  \sum_{m} \left[ 2 J \cos ( q_\lambda + \Phi m +  \theta) \vert q_\lambda,m \rangle \langle q_\lambda,m \vert \right. \nonumber \\
\left.- J_y\left(  \vert q_\lambda , m+1 \rangle \langle  q_\lambda ,m \vert + \text{h.c.} \right)\right], \label{eq:dispersion}
\end{eqnarray}
and where $q_\lambda\!=\!-\pi + 2 \pi l/ N$ and $l \!\in\! \{0,..., N\!-\!1\}$. The role of the overall phase $\theta$ can be most easily seen by turning off the coupling between the two legs $J_y\!=\!0$. In this case, the spectrum consists of two uncoupled dispersions: one for each leg of the ladder. The minimum of the dispersion for $m\!=\!-1/2$ lies at $q_\lambda\!=\! \pi+ \Phi/2  - \theta$, while that for $m\!=\!+1/2$ lies at $q_\lambda\!=\!\pi- \Phi /2  - \theta$ up to multiples of $2 \pi$ such that $q_\lambda\!\in\! [-\pi, \pi[$. The magnetic flux $\Phi$ therefore controls the relative displacement of the two dispersions, while the second phase $\theta$ controls an overall displacement.

As the coupling $J_y$ is turned on, the dispersions of the two legs are coupled and gaps open where the dispersions cross. Provided that the coupling $|J_y|\! \lesssim\! J$, the energy bands away from the gap openings are not strongly affected. As can be seen in Fig.~\ref{fig:2leg}(a), we can use the phase $\theta$ to set the displacement of the energy bands with respect to $q_\lambda\!=\!0$, with important consequences on the dynamics of a wave packet as we now discuss. 

\subsubsection{Chiral motion of a wave packet in the two-leg ladder}\label{section_reduction}

\begin{figure}[!]
\resizebox{0.49\textwidth}{!}{\includegraphics*{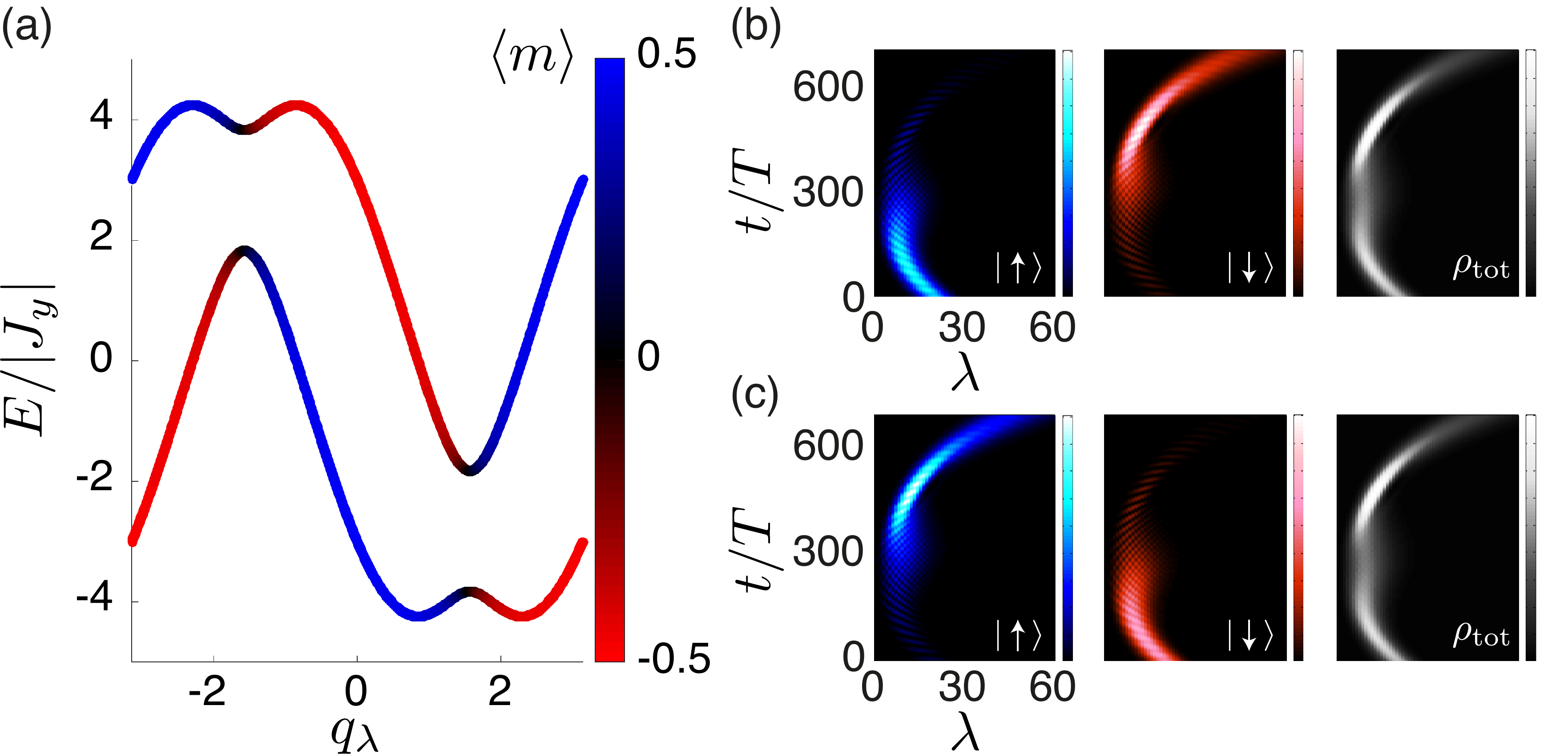}}  
\caption{(a) Spectrum for an isotropic two-leg ladder with \emph{constant} hoppings $J_\lambda \!=\! 2 J_y$ and $\Phi\!=\!\theta\!=\!\pi/2$, where colors specify the mean spin. (b)-(c) Numerical time-evolution using the full Hamiltonian $\hat{H}(t) \!=\!
\hat{H}_0 \!+\! \hat{H}_y \!+\! \hat{V}(t)$. The wave packet is initialised at: (b) $m\!=\!1/2$ and (c) $m\!=\!-1/2$, with $\lambda_0\!=\!20$ and $\sigma\!=\!5$. Spin-resolved and total densities are indicated in the three successive panels. Other parameters are $\Phi\!=\!\theta\!=\!\pi/2$, $\kappa \!=\!0.01 \omega  / l_{\text{H}}$, $J_y\!=\!J_{\lambda_0}/2\!\approx\!8\!\times\! 10^{-3}\omega$, $\Delta\!=\!0$, so as to resemble the isotropic model in (a). Due to the chiral motion, the average energy only increases beyond its initial value after a time $t\approx 550 T$.}
 \label{fig:2leg} 
\end{figure}

We consider a Gaussian wave packet [Eq.~\eqref{eq:gauss}], built from the states making up the energy dispersion in Fig.~\ref{fig:2leg}(a). Such a wave packet is initially centered at $q_\lambda\!=\!0$, and so, for these parameters, is predominantly built from states with a negative group velocity, leading to propagation down the synthetic dimension. Besides, the initial spin $m$ of the wave-packet determines which branch of Fig.~\ref{fig:2leg}(a) is populated. Changing the overall phase $\theta$ shifts the spectrum with respect to $q_\lambda\!=\!0$, as discussed above, and so can be used to control the mean velocity of the wave-packet COM. 

The chiral behaviour of a wave packet persists also with anisotropy, as seen in Figs.~\ref{fig:2leg}(b)-(c) from the full time-dynamics, and as further shown in Appendix~\ref{sec:chiral} from chiral currents. For the full time-dynamics, the wave packet is initialised either in the spin $m\!=\!1/2$ [Fig.~\ref{fig:2leg}(b)] or $m\!=\!-1/2$ [Fig.~\ref{fig:2leg}(c)]. In both cases, the wave packet moves down the synthetic dimension until it hits a ``hard wall" at $\lambda=0$; then particles are transferred to the other spin-state, with opposite velocity, and the wave packet begins at last to move up.  It is remarkable that the corresponding reduction of average energy $E_{\text{av}}(t)$ upon driving can be simply interpreted in terms of the chirality generated by the synthetic flux $\Phi$ in the synthetic-dimension picture.
We note that the total time taken for the average energy $E_{\text{av}}(t)$ to increase past its initial value is $t\!\approx\!550T$ in Fig.~\ref{fig:2leg}(d), in contrast to the immediate increase in $E_{\text{av}}(t)$ in the absence of the artificial magnetic field~[Fig.~\ref{fig:classical}(a)]. Note that if we employ a different overall phase, such as $\theta\!=\!-\pi/2$, wave packets have a positive group velocity and so the average energy increases straight away. We point out that the sensitivity to the initial phase of the drive ($\theta$) can be avoided by adiabatically launching the time-modulation~\cite{Goldman:2014}. As well as the ``hard wall" at $\lambda\!=\!0$, there may also  be a ``soft wall" at high $\lambda$ due to anharmonicity in the harmonic trap.

\begin{figure}[!]
\includegraphics[width=6.5cm]{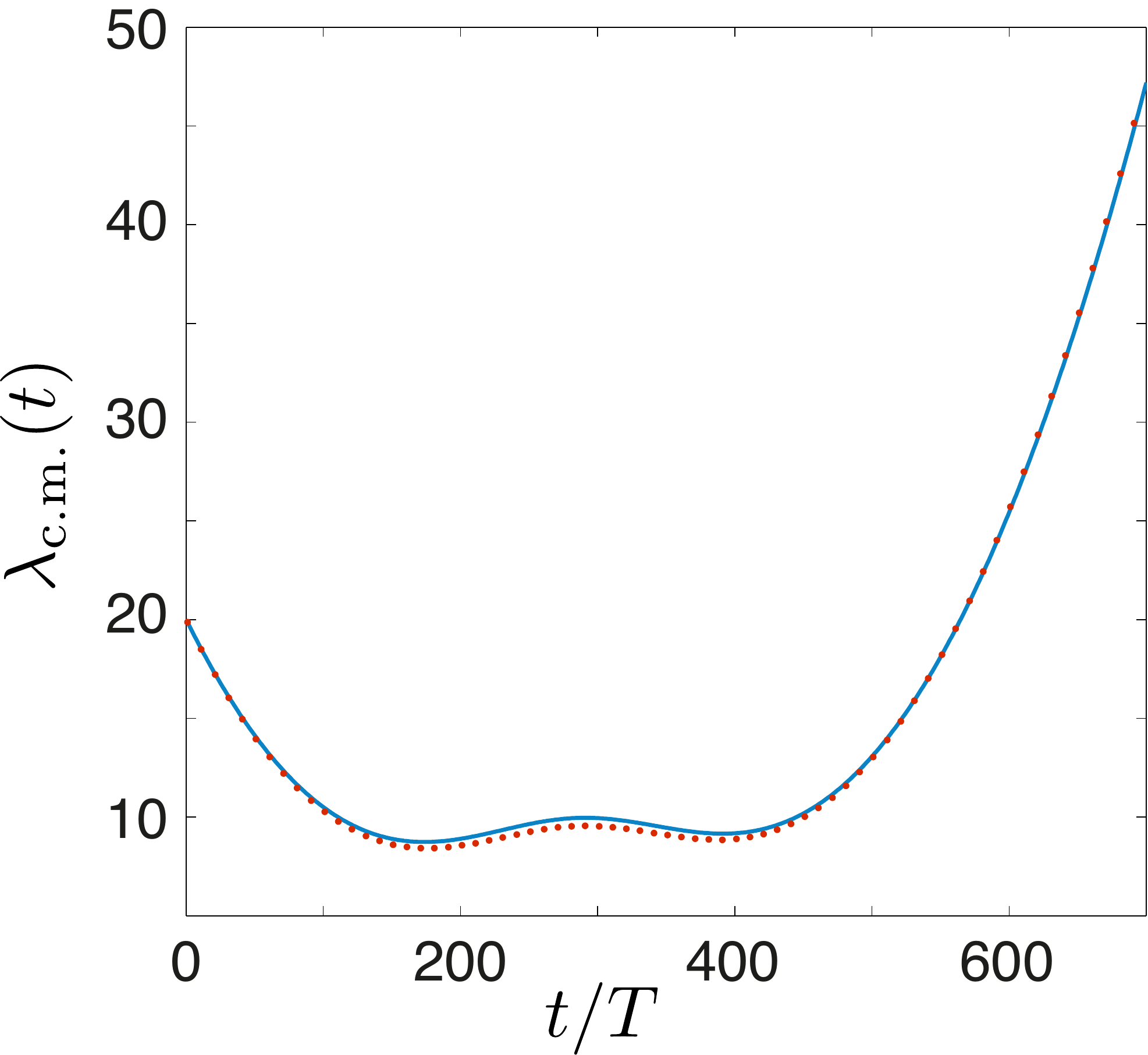}
\vspace{-0.cm} \caption{Center-of-mass trajectory $ \lambda_{\text{c.m.}} (t)$ obtained from a direct numerical evaluation of the time-evolution operator $\hat U (T_{\text{D}};0)$ associated with the full time-dependent Hamiltonian~\eqref{H_tot_2D} (red dots), and compared with the trajectory generated using the effective Hamiltonian~\eqref{effective_ham_2D}. Here the detuning is $\Delta\!=\!0$, hence $T\!=\!T_{\text{D}}\!=\!2 \pi/\omega$, and all other parameters are given in the caption of Fig.~\ref{fig:2leg}(b). Note that in both cases, the time-evolution is probed stroboscopically, at times given by $t\!=\!N T_{\text{D}}$, with $N$ integer [i.e.~dynamics are generated by $[\hat U (T_{\text{D}};0)]^N$, with $N\!=\!1,\dots,700$]. }\label{Fig_eff_num_ladder}\end{figure}

We can also validate the effective Hamiltonian in Eq.~\eqref{effective_ham_2D} quantitatively by analyzing the center-of-mass trajectories associated with the two-leg ladder configuration, i.e.~$m=\pm 1/2$. Specifically, we show in Fig.~\ref{Fig_eff_num_ladder}, the center-of-mass $ \lambda_{\text{c.m.}} (t)$ corresponding to the time-evolving (total) density shown in Fig.~\ref{fig:2leg}(b). The comparison between a direct numerical evaluation of the dynamics associated with the full time-dependent Hamiltonian (red dots) and the effective Hamiltonian prediction (blue line), shows excellent agreement throughout the time evolution. 

\subsection{Reaching the 2D Quantum Hall Regime} \label{sec:2D}

In this section, we increase the number of sites in the second direction, such that particles now move in a (synthetic) 2D lattice in the $\lambda\!-\!m$ plane [Fig.~\ref{fig:2conf}(c)]. As introduced above, this configuration is described by Eq.~\eqref{effective_ham_2D} with $m\!\in\!\mathbb{Z}$. We now choose the modulation phase as $\phi_m= \Phi m $ (i.e. hereafter, we set $\theta\!=\!0$). Then this effective Hamiltonian~\eqref{effective_ham_2D} corresponds to an (anisotropic) Harper-Hofstadter (HH) model~\cite{Hofstadter}, with a uniform flux $\Phi$ per plaquette. This is a seminal lattice model for studying the 2D quantum Hall effect, as it has bulk bands with non-trivial topology that can be associated with robust chiral one-way edge states. As we now show, such chiral edge modes persist even in the presence of the anisotropy in Eq.~\eqref{effective_ham_2D}, and so our scheme for utilising the HO states as a synthetic dimension provides a simple route towards investigating QH physics in a wide-range of ultracold gas set-ups. 

While we focus primarily on edge physics, we also emphasise that our scheme can be used to probe bulk topology, e.g. by measuring the quantised Hall conductivity of a filled band in the presence of a non-zero detuning, $\Delta$, acting as a force along the synthetic dimension.\\

\subsubsection{Energy dispersion of the anisotropic HH model}

We begin by calculating the energy dispersion of the 2D anisotropic HH model in Eq.~\eqref{effective_ham_2D}. Let us start from the Harper equation associated with the Hamiltonian in Eq.~\eqref{effective_ham_2D}, in the absence of detuning $\Delta\!=\!0$. This reads
\begin{align}
&E \psi (\lambda, m)= -J_y \left [ \psi (\lambda, m+1)+\psi (\lambda, m-1)    \right ] \label{schrodinger_one}\\
& + J_{\lambda +1} e^{i \Phi m} \psi (\lambda+1, m)+ J_{\lambda } e^{-i  \Phi m} \psi (\lambda-1, m).\notag
\end{align}
The Hamiltonian matrix associated with Eq.~\eqref{schrodinger_one} explicitly depends on both coordinates $m$ and $\lambda$. In order to investigate the edge modes propagating along the lower edge of the lattice (i.e.~$\lambda\!=\!0$, $m\!=\!1, \dots, L_y$), it is instructive to make a gauge transformation that removes all dependence on the $m$ coordinate. This is achieved through the transformation
\be
\psi (\lambda, m)=\tilde \psi (\lambda, m)e^{-i \Phi m \lambda}, \label{gauge_transf}
\ee
which results in the modified Harper equation
\begin{align}
&E \tilde \psi (\lambda, m)=  J_{\lambda +1}  \tilde \psi (\lambda+1, m)+ J_{\lambda } \tilde  \psi (\lambda-1, m) \label{schrodinger_two}\\
&-J_y \left [ e^{-i  \Phi \lambda}\tilde \psi (\lambda, m+1)+e^{i  \Phi \lambda}\tilde \psi (\lambda, m-1)    \right ] .\notag
\end{align}
The corresponding Hamiltonian matrix only depends on the $\lambda$ coordinate through the anisotropic tunneling parameters $J_\lambda$ and the modified Peierls phase-factors. Now we solve the problem on a cylinder aligned along $\lambda$ (i.e. with periodic boundary conditions applied along $y$ and open boundary conditions along $\lambda$), such that the solution of Eq.~\eqref{schrodinger_two} takes the form
\be
\tilde \psi (\lambda, m)=e^{i q_y y} \chi (\lambda), \quad q_y \in [-\pi/a , \pi/a [ ,
\ee
where $y\!=\!m a$ denotes the position of lattice sites along the auxiliary direction ($a$ denotes the lattice spacing), and where $q_y$ is the corresponding quasi-momentum, we finally obtain the energy spectrum by solving the simplified (Harper) eigenvalue equation
\begin{align}
E \chi (\lambda)&=  J_{\lambda +1}  \chi (\lambda+1)+ J_{\lambda } \chi(\lambda-1) \label{schrodinger_three}\\
&- 2 J_y \cos \left ( \Phi \lambda -q_y a \right ) \chi(\lambda),\qquad \lambda=1, \dots, \lambda_{\text{cut}} , \notag
\end{align}
where we introduced a cut-off $\lambda_{\text{cut}}\!\gg \!\lambda_0$; in this analysis, we indeed assume that the dynamics take place around some state of reference $\lambda_0$, e.g.~$\lambda_0\!=\!0$ if we focus our attention on the lower edge of the lattice. Diagonalizing the Hamiltonian matrix associated with Eq.~\eqref{schrodinger_three}, we obtain the energy spectrum, as shown in Fig.~\ref{fig:2D}(a). Here $\lambda\!\in\![0, 80]$, but only states with $\langle \lambda \rangle\!<\!30$ are shown; this is a good description for our system provided that the population of higher-energy states is negligible. As in the isotropic HH model, we recognize four ``bands" for $\Phi\!=\!\pi/2$, with the middle bands touching at $E\!=\!0$. The outer and middle bands are also connected by topological chiral edge modes that are well-localised on the system boundary (here $\lambda\!=\!0$). 

\subsubsection{Probing 2D QH physics}

\begin{figure}[!]
\resizebox{0.5\textwidth}{!}{\includegraphics*{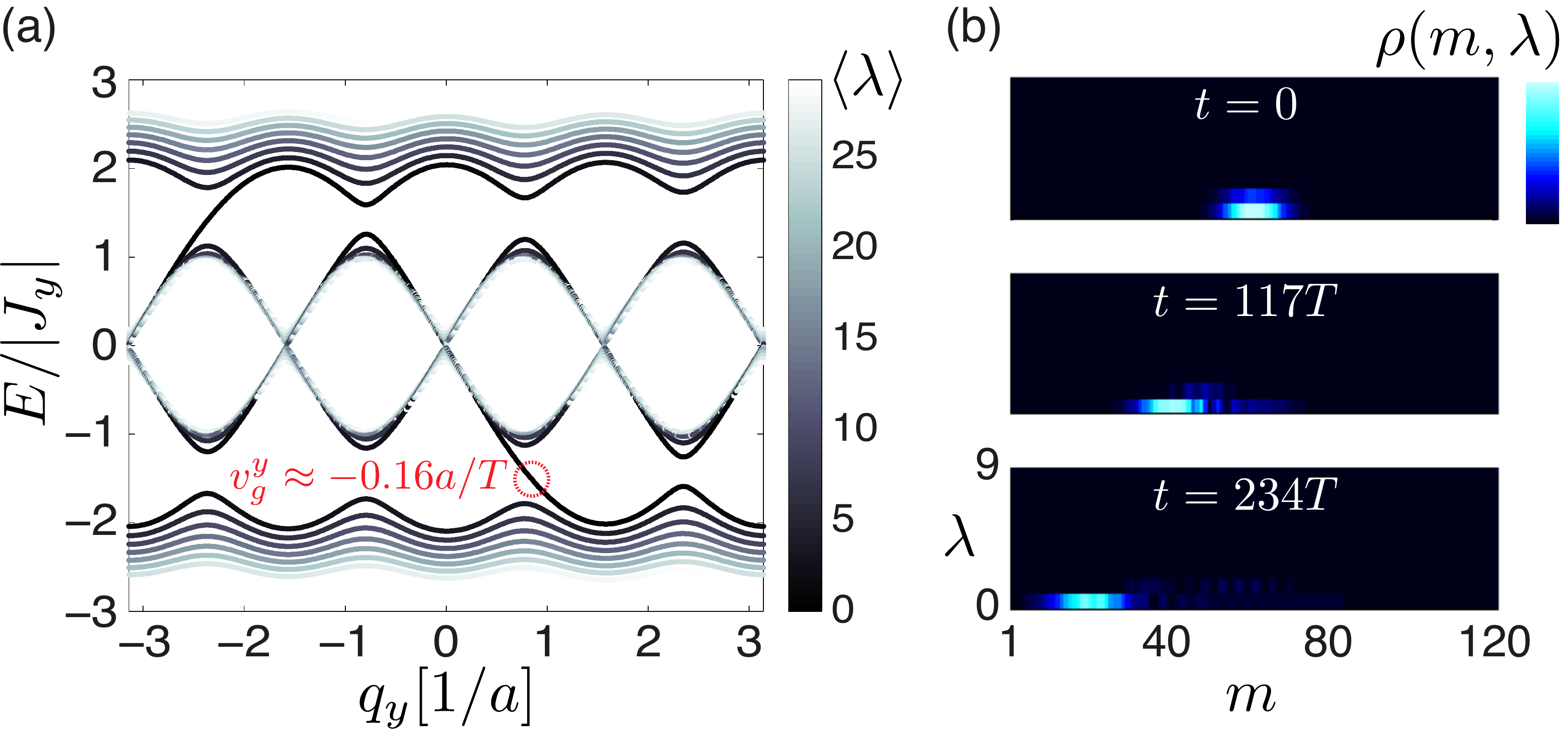}}  
\caption{(a) Spectrum of the 2D anisotropic HH model for periodic boundary conditions along $y$, and $\lambda\!\in\![0, 80]$ with open boundaries; the color scale indicates the weight along $\lambda$, highlighting states with $\langle \lambda \rangle\!<\!30$. The group velocity of the lowest chiral edge mode, at quasi-momentum $q_y\!\approx\!0.8/a$, is indicated. (c) Full time-evolution of a wave packet [Eq.~\eqref{eq:wavepack}], prepared around $m\!\approx\!60$ and $\lambda_0\!\approx\!0$, with mean momentum $\bar q_y\!=\!0.8/a$, and widths given by $\sigma_\lambda\!=\!1$ and $\sigma_y\!=\!7a$; see Eq.~\eqref{eq:wavepack}.  Parameters are $\Phi\!=\!\pi/2$, $\kappa \!=\!0.01 \omega  / l_{\text{H}}$, $J_y\!=\!0.02\omega$ and $\Delta\!=\!0$.}
 \label{fig:2D} 
\end{figure}

To probe the topological chiral edge modes identified in Fig.~\ref{fig:2D}(a), we prepare a Gaussian wave packet so as to efficiently project onto the edge mode localized at $\lambda\!=\!0$. From Fig.~\ref{fig:2D}(a), the group velocity of the edge mode located within the lowest bulk gap is approximatively given by $v_g^y\!=\!-0.16 a /T$, for the realistic system parameters used. We also note that this edge mode is localized in $q_y$-space, around the value $\bar q_y\!\approx\!0.8/a$. This result indicates that a Gaussian wave-packet of the form
\be
\tilde \psi_{\text{gauss}}(\lambda , y) \propto e^{-(y-y_0)^2/2\sigma_y^2} e^{-\lambda^2/2\sigma_\lambda^2} e^{i(y-y_0)\bar q_y},
\ee
will efficiently project onto the edge mode localized at $\lambda\!=\!0$, for $\sigma_{\lambda}$ sufficiently small, and will move chirally along the $-y$ direction. 

Importantly, one should note that the above prediction is only strictly valid when working in the gauge associated with Eq.~\eqref{schrodinger_two}, which is not the gauge realized by the time-dependent Hamiltonian in Eq.~\eqref{H_tot_2D} that leads to the effective Hamiltonian in Eq.~\eqref{effective_ham_2D}. Hence, in the ``real" (experimental) gauge, the wave packet should be prepared in the form 
\be
\psi_{\text{gauss}}(\lambda , y) \propto \tilde \psi_{\text{gauss}}(\lambda , y) e^{- i  \Phi m \lambda},
\ee
where we used the gauge transformation \eqref{gauge_transf}. While creating such a wave packet is challenging from an experimental point of view, we note that its form can be simplified when working in the vicinity of $\lambda_0\!=\!0$, where $e^{- i \Phi m \lambda}\approx 1$. Hence, we conclude that preparing an initial wave packet of the form 
\be
\psi_{\text{gauss}}(\lambda , y) \propto e^{-(y-y_0)^2/2\sigma_y^2} e^{-\lambda^2/2\sigma_\lambda^2} e^{i(y-y_0)\bar q_y}, \label{eq:wavepack}
\ee
should efficiently populate chiral edge modes, for $\sigma_{\lambda}$ sufficiently small. We have validated this prediction in Fig.~\ref{fig:2D}, where we numerically time-evolve the wave packet described by Eq.~\eqref{eq:wavepack}. As shown in Fig.~\ref{fig:2D}(b), there is a clear chiral motion along $y$, in agreement with the dispersion of Fig.~\ref{fig:2D}(a), while the average energy remains constant [$\lambda_{\text{c.m.}}(t)\!\approx\!0$] for all times~(until the wave packet hits a wall along $y$).

As well as edge-state physics, we can also probe the bulk properties of the effective 2D system. In the isotropic HH model, the energy bands can have non-zero topological Chern numbers, leading to QH responses for uniformly-filled bands~\cite{TKNN,COM:2016}. Such quantized responses are robust in the presence of anisotropy provided that the band gap remains open, as shown for 30 sites along $\lambda$ in Fig.~\ref{fig:2D}(a). Under this condition, we can prepare an atomic cloud that uniformly fills the lowest band, e.g.~through Fermi statistics or dephasing effects~\cite{Greiner,Aidelsburger:2015}; energetically, this requires that the temperature is small compared to the band-gap, as quantified in Sec.~\ref{sec:discussion} for realistic experimental parameters. Dynamically, the system can be initialised through appropriate ramping protocols to load the atoms into topological bands, as studied, for example, in Refs.~\cite{Aidelsburger:2015,Kennedy:2015,Hu:2016,Dauphin:2016,Ho:2016}. Once the cloud is prepared, the QH response is then measured as a transverse COM drift~\cite{Dauphin:2013, Aidelsburger:2015,COM:2016} under an applied force. For a force along $\lambda$ (i.e.~$\Delta\!\ne\!0$), this Hall drift will be along the (real) $y$ direction, while for a ``real" force aligned along $y$~\cite{Aidelsburger:2015}, the QH drift will be observed along $\lambda$. In this latter case, the QH response could be used to decrease the average energy  [i.e.~$\lambda_{\text{c.m.}}(t)\!\rightarrow\! 0$], simply by adjusting the orientation of the force (or the flux $\Phi$). Such COM observables could also exhibit non-linear quantized responses~\cite{COM:2016}. 

\section{Using the synthetic dimension to add an {\it extra} dimension } \label{sec:extra}

In the simple implementations sketched in Fig.~\ref{fig:2conf}, we re-interpret dynamics along the HO axis ($x$) in terms of motion along the synthetic ($\lambda$) dimension, while a second dimension is provided either by another real spatial dimension ($y$) or another synthetic dimension, such as the internal atomic states. Now, we introduce a possible generalization of this scheme, which realizes a controllable \emph{extra} dimension: namely, we show how to engineer a synthetic 2D lattice from a 1D lattice structure (i.e.~an array of traps) aligned along a \emph{single} real dimension. This generalized scheme could then be extended so as to create six spatial dimensions out of a 3D atomic system.

In this section, we consider an array of coupled traps aligned along the $x$ direction. A synthetic dimension is then realized within each trap, while coupling between the traps allows one to generate a full (synthetic) 2D lattice.  Below we present two different and complementary schemes, based on optical superlattices and state-dependent potentials respectively, which can be used to enhance and control the inter-trap coupling and so to realise a practical 2D lattice from a single real spatial dimension. We point out that while these two schemes correspond to different physical situations, their theoretical description is very similar.\\

\subsection{Using a time-modulated optical superlattice}

\begin{figure}[h!]
\includegraphics[width=8cm]{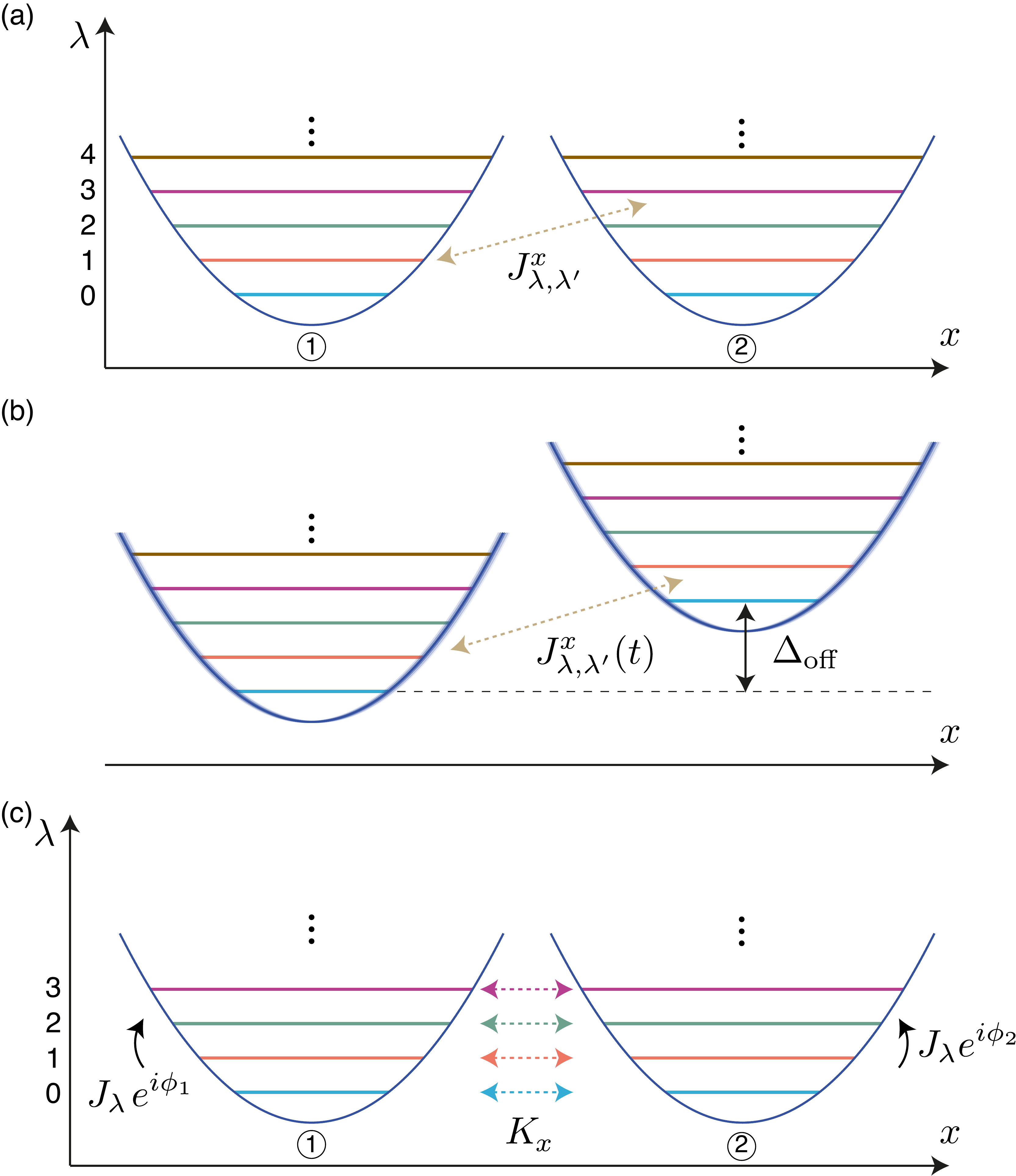}
\vspace{-0.cm} \caption{Generating an extra dimension, using a time-modulated optical superlattice. (a) Two neighboring harmonic wells, denoted 1 and 2, respectively. The coupling matrix elements between harmonic-oscillator eigenstates associated with the two wells are denoted $J_{\lambda, \lambda '}^x$. (b) Introducing an energy offset  $\Delta_{\text{off}}$ between the wells and a time-modulation of the wells depth~\cite{Alberti,Choudhury} allows one to control the hopping between the wells, resulting in the effective Hamiltonian in Eq.~\eqref{eff_ham_superlattice}. (c) The effective Hamiltonian in Eq.~\eqref{eff_ham_superlattice} describes ``diagonal" hopping along $x$ and along $\lambda$.} \label{Fig_Extra_one}\end{figure}

We start by considering two neighboring harmonic traps, both aligned along the $x$ direction, and we assume that this two-well system can be described by a static Hamiltonian of the form
\begin{align}
\hat H_0 = &\omega \sum_{\lambda} \lambda \left (\vert \lambda, 1 \rangle \langle \lambda, 1 \vert + \vert \lambda, 2 \rangle \langle \lambda, 2 \vert \right )\notag \\
&+ \sum_{\lambda, \lambda '} J_{\lambda, \lambda '}^x  \vert \lambda, 1 \rangle \langle \lambda ', 2 \vert + \text{h.c.},\label{static_super}
\end{align}
where $\omega$ is the trap frequency and where $\vert \lambda ,1 \rangle$ [resp.~$\vert \lambda ,2 \rangle$] are the harmonic-oscillator (HO) eigenstates defined at each well; see Fig.~\ref{Fig_Extra_one}(a). The second line in Eq.~\eqref{static_super} describes the coupling between the HO eigenstates defined within each well, and $J_{\lambda, \lambda '}^x$ denotes the corresponding hopping amplitudes (which can be evaluated, e.g., using a tight-binding approach~\cite{Simonbook}). We point out that such couplings can involve different HO levels, $\lambda\,\ne\,\lambda '$; in the following, we will show how additional ingredients can be used to enhance the purely ``diagonal" hopping matrix elements along the $x$ direction ($J_{\lambda, \lambda}^x$), in addition to the induced hopping along $\lambda$ within each trap (as described in Sec.~\ref{sec:syn}). Here, ``diagonal" hopping refers to a situation where the coupling between the wells only involves the same HO levels from either side; this is required for creating a synthetic square lattice in the $x-\lambda$ plane, upon the addition of many neighboring wells. 

 To enhance and control the diagonal hopping between different wells, we introduce two ingredients: a constant energy offset $\Delta_{\text{off}}$ between the neighboring wells, as well as a resonant time-modulation of the lattice depth~\cite{Alberti,Choudhury}. This results in two additional terms in the Hamiltonian,
\begin{align}
\hat H_{\text{res}} = & \Delta_{\text{off}} \sum_{\lambda} \lambda \vert \lambda, 2 \rangle \langle \lambda, 2 \vert \notag \\
&+ 2 \cos ( \Delta_{\text{off}} t) \sum_{\lambda, \lambda '} K_{\lambda, \lambda '}^x  \vert \lambda, 1 \rangle \langle \lambda ', 2 \vert + \text{h.c.},\label{static_super_mod}
\end{align}
as illustrated in Fig.~\ref{Fig_Extra_one}(b). Here, the quantity $K_{\lambda, \lambda '}^x$ parametrizes the modulation of the tunneling matrix elements. Combining all these ingredients, our system is described by the total Hamiltonian $\hat H_{\text{tot}}\,=\,\hat H_0 \!+\! \hat H_{\text{res}} \!+\! \hat V (t)$, where 
\begin{align}
\hat V(t) = 2 \sum_{\lambda}& J_{\lambda} \biggl ( \vert \lambda,1 \rangle \langle \lambda -1, 1 \vert \cos (\omega_{\text{D}} t \!+\! \phi_1)   \notag \\
+& \vert \lambda,2 \rangle \langle \lambda -1, 2 \vert \cos (\omega_{\text{D}} t \!+\! \phi_2)  \biggr) +\text{h.c.},  \label{modulation_basis_bis}
\end{align}
is the time-modulation generating motion along the extra ($\lambda$) dimension; see Eqs.~\eqref{modulation_basis} and \eqref{J_lambda}. Note that the modulation phases $\phi_{1,2}$ could be different within the two different wells (which is a requirement for generating artificial fluxes; see below). In the following, we neglect the detuning $\Delta$ and hence write $\omega_{\text{D}}\!=\!\omega$.

As in Section~\ref{sec:syn}, we find it convenient to rewrite the total Hamiltonian $\hat H_{\text{tot}}$ in a moving frame; in this case, we consider the frame associated with the unitary transformation
 \be
\hat R (t)= \exp \left ( i t \sum_{\lambda}  \left[ \lambda \omega \vert \lambda ,1 \rangle \langle \lambda ,1 \vert +(\lambda \omega +  \Delta_{\text{off}}) \vert \lambda ,2 \rangle \langle \lambda ,2 \vert  \right]\right ), \notag
\ee
which results in the final Hamiltonian
\begin{align}
\hat H_{\text{eff}}=\sum_{\lambda} &J_{\lambda} \biggl (\vert \lambda -1,1 \rangle \langle \lambda, 1 \vert e^{i \phi_1}+ \vert \lambda -1,2 \rangle \langle \lambda, 2 \vert e^{i \phi_2}   \biggr)  \notag \\
+ \sum_{\lambda} &K_x (\lambda)  \, \vert \lambda, 1 \rangle \langle \lambda , 2 \vert \quad + \text{h.c.}\label{eff_ham_superlattice}
\end{align}
where we again performed a RWA by neglecting the fast-rotating terms, and where we introduced the short notation $K_x(\lambda)\!=\!K_{\lambda, \lambda}^x$ for the tunneling matrix element along $x$. For the RWA approximation to be valid, we assumed that the two modulation frequencies satisfy $( \Delta_{\text{off}}, \omega) \gg (J_{\lambda}, K_x)$; we also note that the effective Hamiltonian in Eq.~\eqref{eff_ham_superlattice} is valid even when the frequencies are chosen to be commensurate, $\omega\!=\! p  \Delta_{\text{off}}$, with $p\!\in\!\mathbb{N}$, as long as $p>2$. The validity of the effective Hamiltonian in Eq.~\eqref{eff_ham_superlattice} has been verified numerically, following the procedure detailed in Sec.~\ref{section:validity}. Importantly, this effective Hamiltonian describes ``diagonal" hopping along $x$ and along $\lambda$, as desired and as illustrated in Fig.~\ref{Fig_Extra_one}(c). In particular, the modulation strength $K_x$ can be exploited to tune and enhance the hopping between the traps, while the phases $\phi_{1,2}$ can be chosen so as to generate effective magnetic fluxes in the square plaquettes defined in this $x-\lambda$ plane.

We have shown how two neighboring wells, aligned along a single spatial direction ($x$), can be coupled in a controllable manner, so as to generate a synthetic two-leg ladder defined in the $x-\lambda$ plane [the two legs of the ladder being associated with the location of the two wells in Fig.~\ref{Fig_Extra_one} (a)]. In this way, the synthetic dimension $\lambda$ indeed constitutes an \emph{extra} dimension, the motion along which can be controlled independently with respect to the motion along the (real) dimension $x$. 

The scheme detailed in above can be readily generalized to the case of $N$ wells, aligned along $x$, with an offset $ \Delta_{\text{off}}$ between neighboring wells; this could be performed using a superlattice optical potential or a Wannier-Stark ladder~(see Ref.~\cite{Goldman:2015} and references therein). Such a configuration would then result in a full (synthetic) 2D lattice, defined in the $x-\lambda$ plane, starting from a single spatial dimension ($x$). We point out that the tunneling matrix elements associated with such a synthetic 2D lattice would be non-uniform (and anisotropic), since the matrix elements $K_x (\lambda)$ and $J_{\lambda} $ in Eq.~\eqref{eff_ham_superlattice} typically depend on $\lambda$ in a very different manner. Finally, we stress that creating artificial magnetic fluxes in this synthetic 2D lattice would require to control the modulation phase $\phi$ [see Eq.~\eqref{modulation_basis}] independently within each well (as discussed above for the two-well case).\\

\subsection{Using state-dependent optical lattices and laser-induced tunneling}

In order to offer an alternative route to engineering a controllable extra dimension, we now build on the laser-assisted-tunneling schemes of Refs.~\cite{Jaksch,Gerbier}. Specifically,  we propose to combine state-dependent optical lattices with laser-induced-tunneling methods to tune the hopping amplitude along the real spatial dimension ($x$), as we now describe.

\begin{figure}[h!]
\includegraphics[width=8cm]{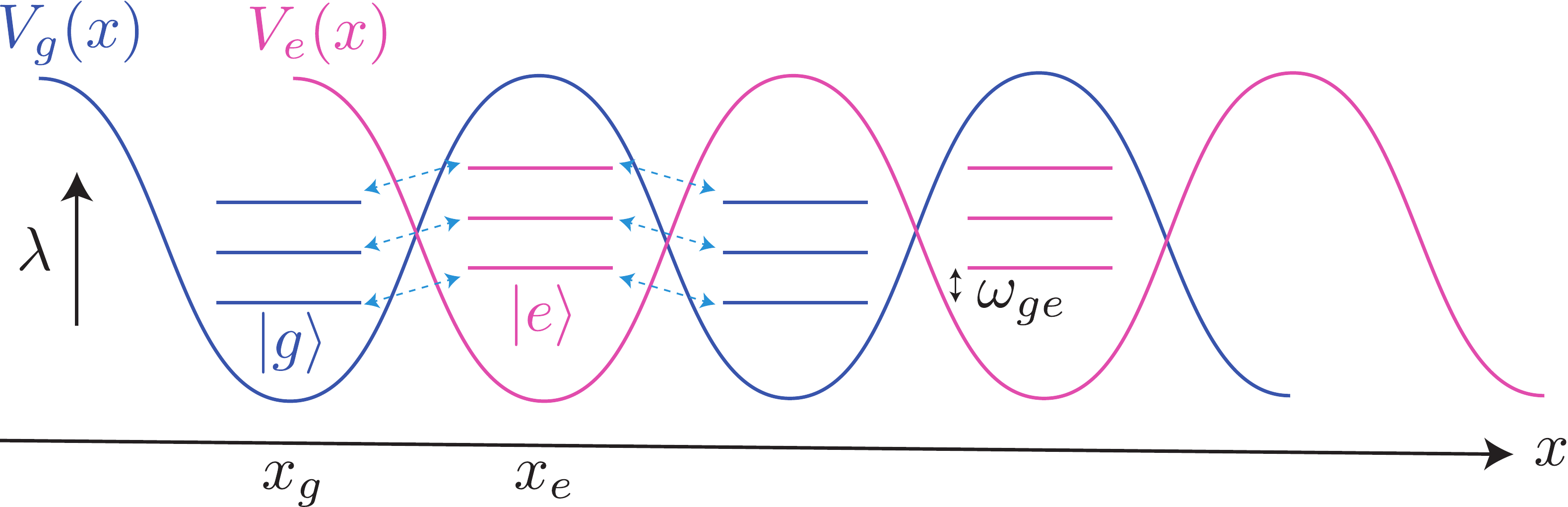}
\vspace{-0.cm} \caption{Generating an extra dimension, using two state-dependent optical lattices.}\label{Fig_Extra}\end{figure}

Consider two internal states of an atom, $\vert g \rangle$ and $\vert e \rangle$, which are respectively trapped in two independent optical lattices $V_{g,e}(x)$, as illustrated in Fig.~\ref{Fig_Extra}. For instance, this could be realized by trapping a ground state ($g$) and a long-lived excited state ($e$) of a two-electron atom (e.g.~ytterbium and alkaline-earth atoms) in an optical lattice whose frequency is set at an ``anti-magic" wavelength~\cite{Gerbier}; alternatively, $g$ and $e$ could represent Zeeman sublevels in the hyperfine ground-state manifold (in which case, an external magnetic field is required to lift the degeneracy). Assuming that the optical-lattice potential is sufficiently strong, tunneling matrix elements are negligible, and the low-energy states associated with each potential well are satisfactorily approximated by harmonic-oscillator (HO) eigenstates. The latter can be coupled within each well, by the time-dependent modulation in Eq.~\eqref{modulation_basis}, so as to generate the synthetic dimension ($\lambda$). In addition, one can introduce a laser field that couples the two internal states, so as to induce effective coupling between the neighboring wells [Fig.~\ref{Fig_Extra}]. 

In order to evaluate the form of this coupling, let us consider two neighboring sites ($x_g$ and $x_e$ in Fig.~\ref{Fig_Extra}). The corresponding (static) Hamiltonian is taken to be of the simple form
\begin{align}
\hat H_0 = &\omega \sum_{\lambda} \lambda \left (\vert \lambda, g \rangle \langle \lambda, g \vert + \vert \lambda, e \rangle \langle \lambda, e \vert \right )\notag \\
&+ \omega_{ge} \sum_{\lambda} \vert \lambda, e \rangle \langle \lambda, e \vert,
\end{align}
where $\omega$ is the trap frequency at the bottom of each well, $\omega_{ge}$ is the atomic transition frequency, and where $\vert \lambda ,g \rangle$ [resp.~$\vert \lambda ,e \rangle$] are the harmonic-oscillator eigenstates defined at the lattice site $x_g$ [resp.~$x_e$]; we write the corresponding wave functions as
\be
\langle x \vert \lambda , g \rangle = \phi_{\lambda} (x - x_g), \qquad \langle x \vert \lambda , e \rangle = \phi_{\lambda} (x - x_e).
\ee
For the sake of simplicity, we assumed that the harmonic-oscillator structures are equivalent in all the potential wells and that the bare tunneling is completely suppressed between these lattice sites; however, we note that these are not crucial assumptions. Next, we introduce an atom-light coupling, which is resonant with the atomic transition frequency~\cite{Dalibard:notes}
\begin{align}
&\hat V_{\text{coupl}} (t) = \Omega \vert g \rangle \langle e \vert \cos (\omega_{ge} t) + \text{h.c.},\label{coupling_atom}\\
&\qquad \quad \, \, \, \,  =\Omega \sum_{\lambda, \tilde \lambda} V_{\lambda \tilde \lambda} \vert \lambda, g \rangle \langle \tilde \lambda, e \vert \cos (\omega_{ge} t ) + \text{h.c.},\notag 
\end{align}
where we introduced  the Rabi frequency $\Omega$ as well as the coefficients
\be
V_{\lambda \tilde \lambda}= \int \text{d}x \, \phi_{\lambda} (x - x_g) \phi_{\tilde \lambda} (x - x_e).
\ee
It is then convenient to represent the total Hamiltonian $\hat H_{\text{tot}}\!=\!\hat H_0\!+\! \hat V_{\text{coupl}} (t)\!+\! \hat V (t)$ in a rotating frame, using the following unitary operator
\be
\hat R (t)= \exp \left ( i t \sum_{\lambda} \lambda \omega \vert \lambda ,g \rangle \langle \lambda ,g \vert +(\lambda \omega + \omega_{ge}) \vert \lambda ,e \rangle \langle \lambda ,e \vert \right ). \notag
\ee
We note that the time-modulation generating the synthetic dimension now reads
\begin{align}
\hat V(t) = 2 \sum_{\lambda} & J_{\lambda} \biggl ( \vert \lambda,g \rangle \langle \lambda -1, g \vert \cos (\omega_{\text{D}} t \!+\! \phi_1)   \notag \\
+& \vert \lambda,e \rangle \langle \lambda -1, e \vert \cos (\omega_{\text{D}} t \!+\! \phi_2)  \biggr) +\text{h.c.},  \label{modulation_basis_tri}
\end{align}
similarly to Eq.~\eqref{modulation_basis_bis}.
After applying the RWA, we eventually find that the two-site system is well captured by the effective Hamiltonian
\begin{align}
&\hat H_{\text{eff}} \approx \sum_{\lambda} J_{\lambda} \biggl (\vert \lambda -1,g \rangle \langle \lambda, g \vert e^{i \phi_1}+ \vert \lambda -1,e \rangle \langle \lambda, e \vert e^{i \phi_2}   \biggr) \notag\\
&\qquad \qquad +J_x^{(\lambda)} \vert \lambda , g \rangle \langle \lambda ,e \vert + \text{h.c.},\\
&J_x^{(\lambda)}=\frac{\Omega}{2}  \int \text{d}x \, \phi_{\lambda} (x - x_g) \phi_{\lambda} (x - x_e).
\end{align}
Note that the effective tunneling matrix element $J_x^{(\lambda)}$ is real, however, it could be made complex by introducing a phase in the atom-light coupling [Eq.~\eqref{coupling_atom}]; this could be used to engineer additional gauge structures in the 2D synthetic lattice spanned by the (real) lattice sites, and the synthetic dimension. We also point out that the tunneling matrix elements $J_x^{(\lambda)}$ typically depend on $\lambda$, which produces anisotropy in the synthetic 2D lattice.

Considering the many sites of the state-dependent lattices leads to an effective Hamiltonian of the form,
\begin{align}
\hat H_{\text{eff}} \approx & \sum_{\lambda , n} J_x^{(\lambda)} \biggl ( \vert \lambda , n \rangle \langle \lambda , n+1 \vert + \text{h.c.}\biggr ), \notag\\
&+J_{\lambda} \biggl ( \vert \lambda -1 ,n \rangle \langle \lambda ,n \vert e^{i \phi_n}  + \text{h.c.} \biggr )
\end{align}
where we introduced the site index $n\!=\!x/a$, with the unit length $a\!=\!x_e\!-\!x_g$.  We note that to engineer non-trivial fluxes in this synthetic 2D lattice, the modulation phases $\phi_n$ should depend on the site index $n$, which indicates that the time-modulation generating the synthetic dimension should be applied locally (with a single-site resolution).

\section{Interactions in the synthetic dimension} \label{sec:inter}

We now discuss interactions along the synthetic dimension. We assume that, as in ultracold atomic gases, the interaction is zero-range in terms of the original coordinate $x$, with a Hamiltonian given by
\begin{align}
	\hat{H}_\mathrm{int}
	=
	\frac{U_0}{2}
	\int dx
	\hat{\psi}^\dagger (x) \hat{\psi}^\dagger (x) \hat{\psi} (x) \hat{\psi} (x),\label{contact}
\end{align}
where $U_0$ is the interaction strength and $\hat{\psi}(x)$ is the annihilation operator of a particle at position $x$.
We can expand the field operator in terms of harmonic oscillator eigenstates as
\begin{align}
	\hat{\psi}(x)
	&=
	\sum_\lambda \langle x | \lambda \rangle \hat{a}_\lambda
	\notag \\
	&=
	\sum_\lambda
	\sqrt{\frac{1}{ l_{\text{H}}  2^\lambda \lambda! \sqrt{\pi  } }}
	H_\lambda (x/l_{\text{H}})e^{- x^2/ (2 l^2_{\text{H}})}\hat{a}_\lambda
	\notag \\
	&\equiv
	\frac{1}{\sqrt{l_{\text{H}}}}
	\sum_\lambda
	h_\lambda (\xi) \hat{a}_\lambda,
\end{align}
where $\hat{a}^\dagger_\lambda$ creates a particle in the state $|\lambda\rangle$ in the laboratory (non-rotating) frame, $H_\lambda (\xi)$ is the Hermite polynomial, and we have introduced a dimensionless variable $\xi\! \equiv\! x/l_{\text{H}}$. We have also defined normalized Hermite polynomials as
\begin{align}
	h_\lambda (\xi) \equiv \frac{1}{\sqrt{2^\lambda \lambda! \sqrt{\pi}}}
	H_\lambda (\xi) e^{-\xi^2/2}.
\end{align}
The interaction Hamiltonian can then be written as
\begin{align}
	\hat H_\mathrm{int}
	=
	\frac{U_0}{2l_{\text{H}}}\sum_{\lambda_1,\lambda_2,\lambda_3,\lambda_4}\int d\xi&
	h_{\lambda_1} (\xi) h_{\lambda_2} (\xi) h_{\lambda_3} (\xi) h_{\lambda_4} (\xi)
	\notag \\ 
	&\times \hat{a}^\dagger_{\lambda_4} \hat{a}^\dagger_{\lambda_3} \hat{a}_{\lambda_2} \hat{a}_{\lambda_1}.
\end{align}
In general, there is always a nonzero interaction among states with any combination of $\lambda_1$, $\lambda_2$, $\lambda_3$, and $\lambda_4$, provided that $\lambda_1 + \lambda_2 + \lambda_3 + \lambda_4$ is even.
However, we can restrict this sum further by using the rotating wave approximation, which here requires that $\omega \gg U_0$.
Going to the rotating frame by $\hat{c}_\lambda \equiv \hat{a}_\lambda e^{i\lambda \omega t}$ and ignoring fast-oscillating terms, we obtain
\begin{align}
	&\hat H_\mathrm{int}
	=
\frac{U_0}{2l_{\text{H}}}
	\sum_{\lambda_1 + \lambda_2 = \lambda_3 + \lambda_4}
	U(\lambda_1,\lambda_2;\lambda_3,\lambda_4)
	\hat{c}^\dagger_{\lambda_4} \hat{c}^\dagger_{\lambda_3} \hat{c}_{\lambda_2} \hat{c}_{\lambda_1},
\end{align}
where
\begin{align}
	U(\lambda_1,\lambda_2;\lambda_3,\lambda_4)
	\equiv
	\int d\xi
	h_{\lambda_1} (\xi) h_{\lambda_2} (\xi) h_{\lambda_3} (\xi) h_{\lambda_4} (\xi),
\end{align}
and where we emphasise that the sum over $\lambda$ is now restricted to processes which conserve $\lambda$, namely $\lambda_1 + \lambda_2\! =\! \lambda_3 + \lambda_4$.
Physically, the quantity $U(\lambda_1,\lambda_2; \lambda_3,\lambda_4)$ characterizes the relative interaction strength for a process in which particles in states $\lambda_1$ and $\lambda_2$ collide and scatter into states $\lambda_3$ and $\lambda_4$. 

While a semi-analytical expression for $U(\lambda_1,\lambda_2;\lambda_3,\lambda_4)$ is available in the mathematical literature~\cite{analytical1, analytical2}, this expression is in general very complicated. We therefore now numerically calculate $U(\lambda_1,\lambda_2;\lambda_3,\lambda_4)$ for some representative cases in order to investigate the structure of the interactions.

\subsection{Numerical estimates of $U(\lambda_1,\lambda_2;\lambda_3,\lambda_4)$}

\begin{figure}[tbp]
\begin{center}
\includegraphics[width=8.5cm]{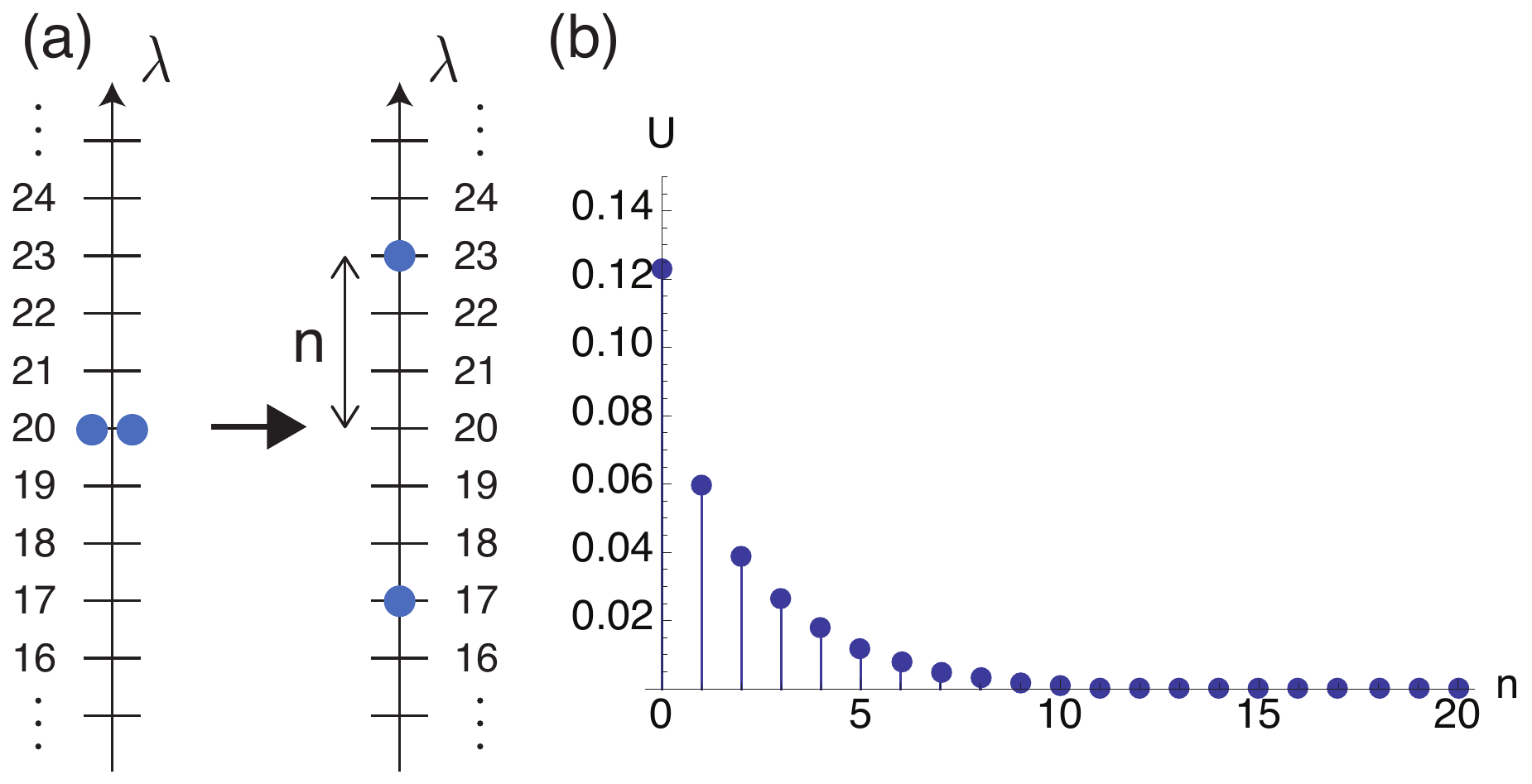}
\caption{(a) Schematic of two particles at $\lambda_1\! =\! \lambda_2 \!= \!20$ scattering into different states $\lambda_3\! = \!20 + n$ and $\lambda_4 \!= \!20 - n$. (b) The corresponding relative interaction strength $U(\lambda_1,\lambda_2;\lambda_3,\lambda_4)$ as a function of $n$.}
\label{firstcase}
\end{center}
\end{figure}

\begin{figure}[tbp]
\begin{center}
\includegraphics[width=8.5cm]{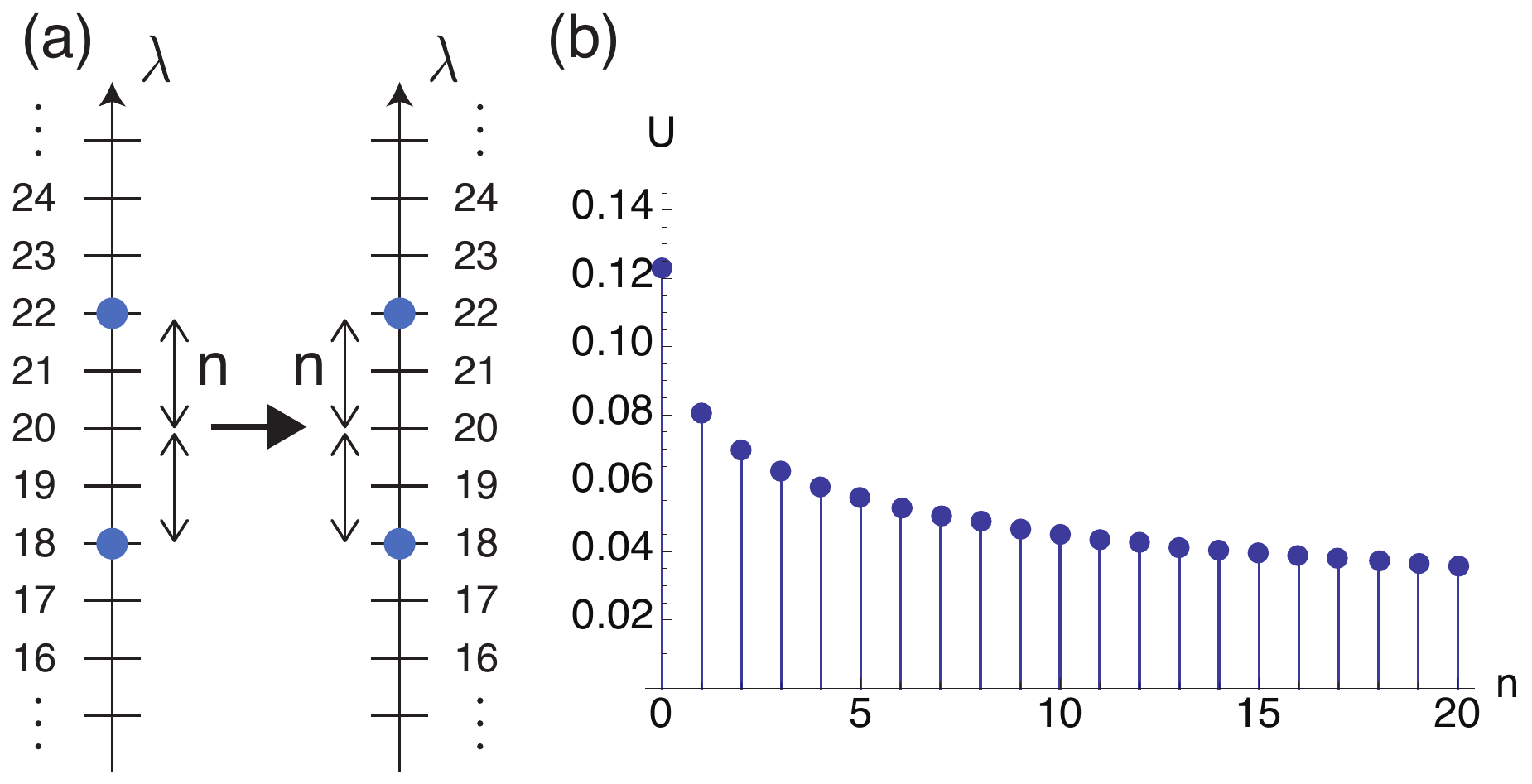}
\caption{(a) Schematic of two particles at $(\lambda_1, \lambda_2) \!=\! (20+n, 20-n)$ scattering into the same states $(\lambda_3, \lambda_4) \!=\! ( 20 + n, 20-n)$. (b) The corresponding relative interaction strength $U(\lambda_1,\lambda_2;\lambda_3,\lambda_4)$ as a function of $n$.}
\label{secondcase}
\end{center}
\end{figure}

\begin{figure}[tbp]
\begin{center}
\includegraphics[width=8.5cm]{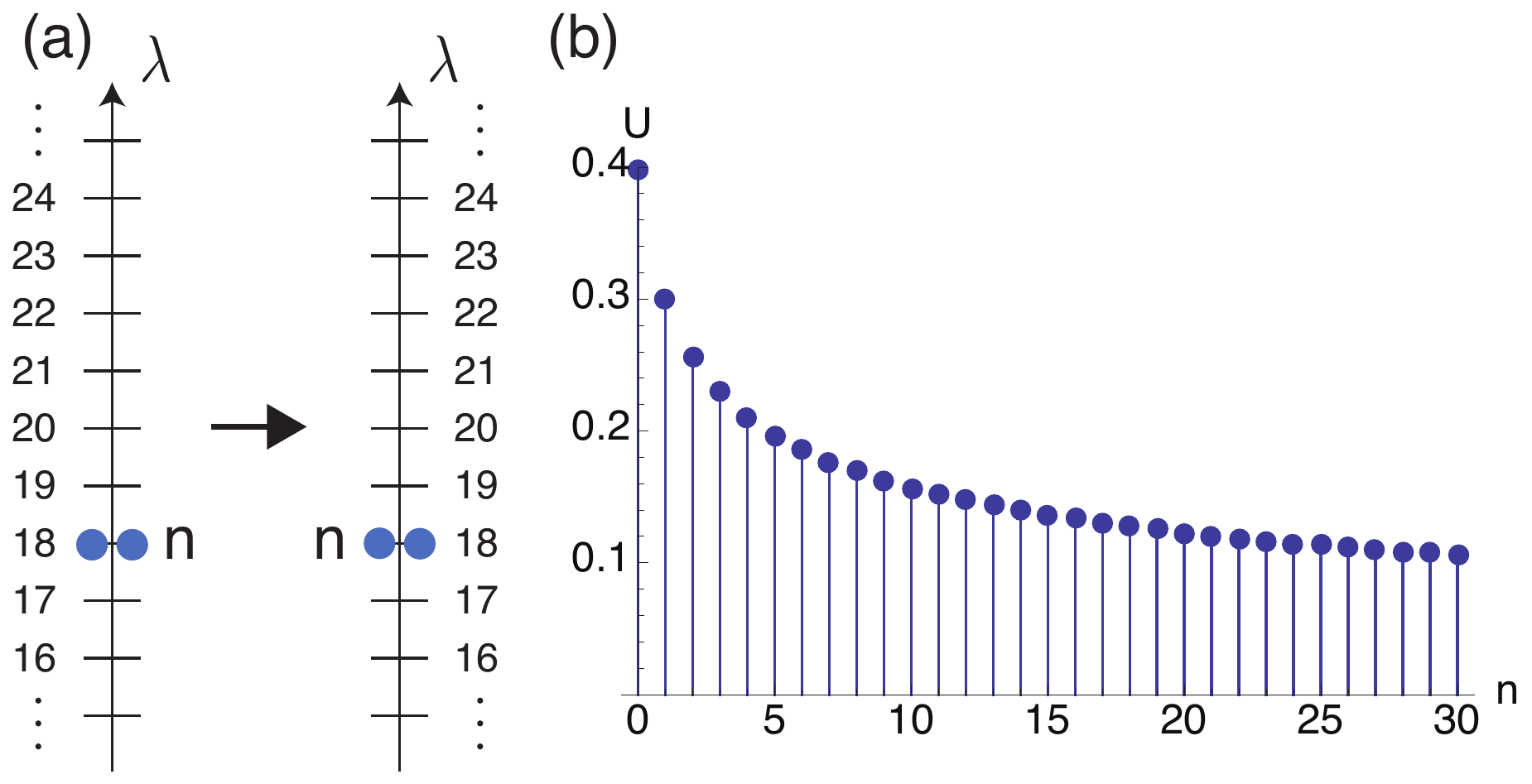}
\caption{(a) Schematic of two-particle scattering when  $\lambda_1 = \lambda_2 = \lambda_3 = \lambda_4 = n$. (b) The corresponding relative interaction strength $U(\lambda_1,\lambda_2;\lambda_3,\lambda_4)$ as a function of $n$.}
\label{thirdcase}
\end{center}
\end{figure}

We first consider the case in which two particles in states $\lambda_1 \!=\! \lambda_2$ scatter into different states $\lambda_3$ and $\lambda_4$.  As is sketched in Fig.~\ref{firstcase}(a), these processes correspond to interaction matrix elements which are off-diagonal in $\lambda$ space. In Fig.~\ref{firstcase}(b), we then plot $U(\lambda_1,\lambda_2;\lambda_3,\lambda_4)$ for $\lambda_1 \!= \!\lambda_2 \!=\! 20$ and $\lambda_3 = 20 + n$ and $\lambda_4\! =\! 20-n$, as a function of $n$, and we observe an exponential drop-off of the relative interaction strength as one increases $n$.
Secondly, we consider two particles in states $(\lambda_1, \lambda_2)$ scattering into the same state $(\lambda_1, \lambda_2)$. These are processes which are diagonal in $\lambda$ space, as shown in Fig.~\ref{secondcase}(a). When we plot the relative interaction strength in Fig.~\ref{secondcase} (b), we now observe that the interaction slowly decays algebraically as the separation between the two states $\lambda_1 - \lambda_2 \!= \!2n$ increases.  We point out that such an algebraic decay of the interaction strength was already analyzed in Ref.~\cite{Swallows}.
Finally, we consider the case in which $\lambda_1\! =\! \lambda_2\! =\! \lambda_3 \!=\! \lambda_4$, as is sketched in Fig.~\ref{thirdcase}(a). Plotting the relative interaction strength in Fig.~\ref{thirdcase}(b), we again observe that the interaction is strongest when $\lambda$ is small and decays slowly as $\lambda$ increases. 

To summarize, we find that the interaction matrix elements which are off-diagonal in $\lambda$ decay exponentially, whereas those that are diagonal in $\lambda$ decay algebraically. The interaction in the synthetic dimension therefore has a  long-ranged nature, and is unlike other typical interactions in ultracold gases and photonics as we now briefly discuss.\\

\subsubsection{ Comparison with on-site interactions along real lattice dimensions}
 For on-site interactions in a lattice, two particles only interact when they are physically on the same lattice site, as there is otherwise a negligible spatial overlap between states. In terms of the notation introduced above, this would correspond to a case in which processes with $\lambda_1\! =\! \lambda_2 \!=\! \lambda_3 \!= \!\lambda_4$ have a constant interaction strength, while all other processes are negligible. In contrast, our interaction matrix elements $U(\lambda_1,\lambda_2;\lambda_3,\lambda_4)$ have a much broader distribution because there can be large non-zero spatial overlaps between different harmonic oscillator states centered around the same position. In terms of the resulting lattice model, interactions are therefore much more ``long-ranged" along the synthetic lattice dimension than on-site interactions are along real lattice dimensions.

\subsubsection{Comparison with SU(N) interactions}
 In the previous implementation of a synthetic dimension in ultracold gases~\cite{Boada2012, Celi:2014, Stuhl:2015, Mancini:2015,Lacki:2016,Barbarino:2015,Zeng:2015,Barbarino:2016,cooper:2016}, different internal atomic states play the role of sites along the synthetic dimension and, to a good approximation, the interactions are SU(N)-invariant, where $N$ is the number of internal states coupled together. Such interactions have two main characteristics: firstly, the internal atomic states cannot be changed in the scattering process, and secondly, each atom interacts equally strongly with all other atoms on the same real-space lattice site, regardless of their respective internal states.  In terms of the synthetic dimension, this second characteristic means that the interactions are effectively zero-dimensional:~an atom on a given lattice site interacts equally strongly with all other atoms regardless of where they are located along the synthetic dimension. This is dramatically different to usual inter-particle interactions in real spatial dimensions, where the interaction strength decays with distance.  We note that the effective zero-dimensional interactions associated with Raman-induced synthetic dimensions may preclude the realisation of fractional QH states~\cite{Lacki:2016}, which are genuinely two-dimensional. 

In the notation introduced above, these SU(N) interactions corresponds to processes with $\lambda_1\! =\! \lambda_4$ and $\lambda_2\! = \!\lambda_3$ having nonzero and constant matrix elements and with all the other processes having zero matrix elements. As demonstrated in Fig.~\ref{secondcase}, our interaction matrix elements have a ``shorter-ranged" structure, as the interaction is stronger when $\lambda_1\! =\! \lambda_4$ and $\lambda_2 \!=\! \lambda_3$ are closer together. This new structure raises the possibility of finding fractional QH states  in our model, that were not possible for the previous implementation of synthetic dimensions using internal atomic states coupled by Raman lasers.

\subsubsection{ Comparison with on-site interactions in Fourier space}
 For integrated photonics, an implementation of synthetic dimensions has been proposed, which harnesses the different modes of a ring resonator as the lattice sites along the synthetic dimension~\cite{4Dphotons:2015}. In this implementation, the interactions are similar to SU(N) interactions, in that each photon interacts equally strongly with all other photons regardless of which modes they occupy, but now the mode indices can be changed in scattering processes. 
In terms of the notation introduced above, these interactions correspond to a case in which the interaction matrix elements are constant irrespective of the values of the four $\lambda$'s involved in the processes. Another way to view this type of photonic interactions is to recognise that they are on-site in the Fourier space of the synthetic dimension~\cite{OzawaCarusotto}. Again, this is different to our interaction matrix elements, which are not on-site in Fourier space and do decay with particle separation along the synthetic dimension.  \\

 We also note that we have performed a similar analysis in which we began from dipolar interactions~\cite{dipolar}, instead of contact interactions [Eq.~\ref{contact}], and we observed no qualitative difference with the results presented here.\\
 
 Finally, we conclude this Section on the effects of interactions by noting that periodically-driven atomic systems can be affected by heating and losses~\cite{Eckardt:Review}, through a complicated interplay between the driving scheme and the interactions. In Bose gases, these effects can lead to a rapid decay of the condensed fraction, which can be traced back to the existence of parametric instabilities at short times; the corresponding instability rates and stability diagrams can be estimated using the method of Ref.~\cite{Lellouch:Parametric}. Moreover, in both Fermi and Bose systems, the heating and losses that arise at longer times can be attributed to two-body scattering processes; the associated heating rates are captured by the so-called Floquet-Fermi-Golden-Rule~\cite{ThomasNigel}.

\section{Discussion}\label{sec:discussion}

In this section, we conclude with some practical remarks, firstly, about the effects on anharmonicity in a typical experimental harmonic trap for cold atoms, and, secondly, concerning possible experimental parameters for a feasible implementation of this scheme. 

\subsection{Comments on Anharmonicity} 

In a cold-atom experiment, a strong harmonic confinement can be created through a (red-detuned) dipole trap. In this case, the dipole potential felt by the atoms takes the form~\cite{Grimm}
\be
U_{\text{dip}} (x) = - \alpha I (x), \quad \alpha >0 ,
\ee
where the real coefficient $\alpha$ is directly related to the polarizability,  $I (x)$ is the intensity of the laser beam, and where we explicitly considered a single spatial direction ($x$) for simplicity. Here, we present, as an example, the case of a focused-beam trap, generated by a single Gaussian laser. Then, the beam intensity is of the form
\be
I(x)= I_0 e^{-\frac{2 x^2}{W^2}} \approx I_0 \left ( 1 - \frac{2 x^2}{W^2} + \frac{2 x^4}{W^4}   \right ),
\ee
where $W$ denotes the waist of the Gaussian beam. The resulting dipole potential is given by
\be
U_{\text{dip}} (x) \approx \left (\frac{2 \alpha I_0}{W^2} \right )x^2 - \left (\frac{2 \alpha I_0}{W^4} \right )x^4, \label{dipole_trap}
\ee
up to an irrelevant constant. Hence, for sufficiently large laser intensity $I_0$ and waist $W$, the potential felt by the atoms is well approximated by a harmonic potential, with the identification
\be
\hat V_{\text{trap}} (\hat x) = \frac{1}{2} M \omega^2 \hat{x}^2 = \left (\frac{2 \alpha I_0}{W^2} \right ) \hat{x}^2 ,
\ee
In order to estimate the effects of anharmonicity in our synthetic-dimension scheme, which is mainly due to the quartic correction in Eq.~\eqref{dipole_trap}, we write the corresponding operator in the harmonic-oscillator-eigenstates basis ($\{ \vert \lambda \rangle \}$). This perturbing effect gives two contributions to the total Hamiltonian in Eq.~\eqref{transf_ham}:  (i)~rapidly-oscillating off-diagonal terms (which vanish in the RWA), and (ii)~a static diagonal term given by
\be
\hat V_{\text{pert}}  = - \sum_{\lambda} \left ( \frac{\sqrt{3}}{2 W \sqrt{M}}  \right )^2 \, \lambda^2 \vert \lambda \rangle \langle \lambda \vert . \label{eq:per}
\ee
where we have omitted a linear term in $\lambda$ which can easily be eliminated through a proper choice of the detuning $\Delta$ [see Eq.~\eqref{transf_ham}]. 
The perturbing term in Eq.~\eqref{eq:per}, which contributes to the effective Hamiltonian in Eq.~\eqref{effective_ham}, corresponds to a residual harmonic potential along the synthetic dimension, 
\be
V_{\text{pert}} (\lambda)= - \frac{1}{2} M \omega_{\lambda}^2 \, \lambda^2 ,
\ee
with effective ``frequency"
\be
\omega_{\lambda}= \sqrt{\frac{3}{2}} \frac{1}{M W},
\ee
which is thus directly set by the waist $W$ of the laser beam. In an experiment, this will limit the ``length" $L_\lambda$ of the synthetic dimension over which our effective models well-describes the dynamics; practically, it is expected that $L_{\lambda}\!\sim\!20$ is possible with reasonable laser power~\cite{JPBrantut}.

\subsection{Experimental Parameters}

In Section~\ref{sec:2D}, we found that the QH dynamics observed in our simple model are associated with a time scale $t_{\text{exp}}\!\sim\! 100 T$, which in a cold-atom experiment should be on the order of 10-100ms. This suggests working with a tight harmonic trap of frequency $\omega\!\sim\!10\!-\!100$kHz, and with realistic hopping amplitudes $J_{\lambda,y}\sim0.1\!-\!1$kHz; the corresponding temperature required to resolve the topological gaps is $T_{\text{exp}}\!\lesssim\!1\!-\!10$nK. As the typical group velocity associated with the chiral modes in Fig.~\ref{fig:2D} is $v_g\sim a J_y/\hbar$, a COM displacement would be on the order of 10 lattice sites, which could be directly detected \emph{in situ}~\cite{Aidelsburger:2015}. The driving frequency $\omega_{\text{D}}\!\sim\!10\!-\!100$kHz is also in the acoustic domain, which is experimentally practical.

We also note that to achieve an extra dimension, as proposed in Section~\ref{sec:extra}, laser-assisted tunneling amplitudes of the order of $J_x^{(\lambda)}\!\sim\!$100 Hz could be achieved in schemes involving state-dependent lattices~\cite{Gerbier}. Similar amplitudes have also been experimentally-realized in time-modulated superlattices~\cite{Aidelsburger:2015}. Hence, by adjusting the hopping amplitude associated with the synthetic dimension $J_{\lambda}\!\sim\!J_x^{(\lambda)}$, one should be able to create topological gaps of the order of $\Delta/k_B\!\sim\!$ 10nK, which can be resolved in current experiments.

\acknowledgements The authors acknowledge  A. Bermudez, J.~C. Budich, J.~P. Brantut, I. Carusotto, N. R. Cooper, A. Daley, J. Dalibard, G. Juzeli\=unas, F. Gerbier, M. Oberthaler, T. Scaffidi and F. Schmidt-Kaler for stimulating discussions. H.M.P. and T.O. are supported by the EU-FET Proactive grant AQuS,
Project No. 640800, and by the Autonomous Province
of Trento, partially through the project ``On silicon chip
quantum optics for quantum computing and secure communications"
(``SiQuro"). H.M.P was also supported by
the EC through the H2020 Marie Sklodowska-Curie Action,
Individual Fellowship Grant No: 656093 ``SynOptic". 
N.G. is financed by the FRS-FNRS Belgium and by
the BSPO under PAI Project No. P7/18 DYGEST. 

\appendix
\section{Chiral currents in the anisotopic two-leg ladder}~\label{sec:chiral}

In this Appendix, we further explore the effects of anisotropy on a two-leg ladder in a uniform magnetic field by studying chiral currents along the two legs. As already suggested by the numerical results shown in Fig.~\ref{fig:2leg} in the main text, the anisotropy present in Eq.~\eqref{effective_ham_2D} does not destroy the expected propagation of the wave packet; to further confirm this chiral behaviour, we introduce the current operators associated with the two legs:
\begin{align}
\hat{j}_{\uparrow} &= - i \sum_{\lambda} J_{\lambda +1} \left( e^{-i \left( \frac{\Phi}{2} + \theta\right) } \hat{c}^\dagger_{\lambda+1,\uparrow}\hat{c}_{\lambda,\uparrow} - \text{h.c.} \right) , \notag\\
\hat{j}_{\downarrow} &= - i \sum_{\lambda} J_{\lambda +1} \left( e^{-i \left( - \frac{\Phi}{2} + \theta\right) } \hat{c}^\dagger_{\lambda+1,\downarrow}\hat{c}_{\lambda,\downarrow} - \text{h.c.} \right) ,\label{current_op}
 \end{align}
 where we explicitly used the notations $(\uparrow,\downarrow)$ to designate $m\!=\!\pm 1/2$ and where we have introduced the operators $\hat{c}^\dagger_{\lambda,\uparrow}$ which creates, for example, a particle in the state $\vert \lambda, 1/2\rangle$ in the rotating frame. We then diagonalize the effective Hamiltonian in Eq.~\eqref{effective_ham_2D}, using a system of size $L_{\lambda}$ along the synthetic dimension, to obtain its eigenstates $\{ \vert \chi_n \rangle \}$ and eigenergies $\{ E_n \}$, where $n\!=\!1,\dots, 2L_{\lambda}$. In order to measure the contribution of each state $\vert \chi_n \rangle$ to the currents defined along each leg [Eq.~\eqref{current_op}], we compute the expectation values
 \begin{align}
& \langle \hat{j}_{\uparrow} \rangle_n \!=\! -i \sum_{\lambda} J_{\lambda +1}\! \left( e^{-i \left( \frac{\Phi}{2} + \theta \right )}  \chi_n^*(\lambda+1,\uparrow)\chi_n(\lambda,\uparrow) \!-\! \text{c.c.}   \right),\notag \\
& \langle \hat{j}_{\downarrow} \rangle_n \!=\! -i \sum_{\lambda} J_{\lambda +1}\! \left( e^{-i \left(- \frac{\Phi}{2} + \theta \right )}  \chi_n^*(\lambda+1,\downarrow)\chi_n(\lambda,\downarrow) \!-\! \text{c.c.}   \right),\label{mean_current}
 \end{align}
 where $\chi_n(\lambda,m)\!=\!\langle \lambda,m \vert \chi_n\rangle$. While the sum of the currents along both legs always vanishes, $\langle \hat{j}_{\uparrow} \rangle_n\!+\!\langle \hat{j}_{\downarrow} \rangle_n\!=\!0$, the presence of a sizeable current along a given leg is an important signature of the artificial magnetic flux $\Phi$.  
 
  \begin{figure}[!]
\resizebox{0.47\textwidth}{!}{\includegraphics*{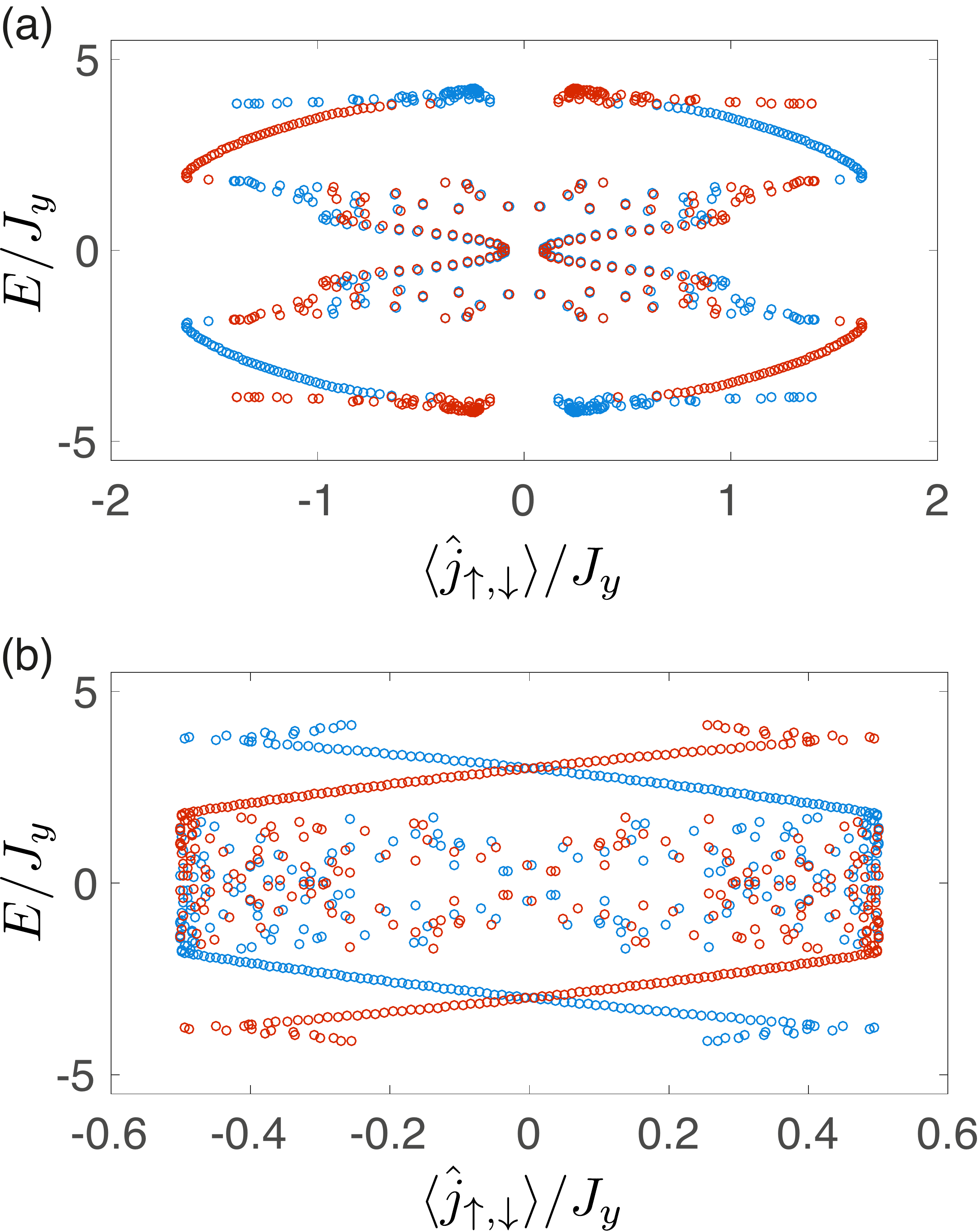}}
\caption{ The current in each leg of the ladder, calculated from the eigenstates of the effective Hamiltonian in Eq.~\eqref{effective_ham_2D}, using Eq.~\eqref{mean_current}. Results are shown for (a) an isotropic two-leg ladder (i.e.~setting $J_{\lambda}\!=\!J\!=\!1$), and for (b) the anisotropic two-leg ladder [Eq.~\eqref{effective_ham_2D}]; blue and red circles correspond to $ \langle \hat{j}_{\uparrow} \rangle_n$ and $ \langle \hat{j}_{\downarrow} \rangle_n$, respectively [see Eqs.~\eqref{current_op}-\eqref{mean_current}]. In each case, we have diagonalised the Hamiltonian for a region of 201 lattice sites along each leg, corresponding in the anisotropic model to $\lambda\! \in\! \{0,..., 200\}$. We also take $\Phi\!=\! \pi/2$ and $\theta\!=\!0$ and all quantities are expressed in terms of the energy scale $J_y$, which is set to $J_y=J/2$ in the isotropic limit and $J_y \!=\! J_{\lambda\!=\!200}/2$ in the anisotropic case. Note how the anisotropy distorts the structure in (a), but still presents energy-dependent chiral behavior in reasonably large energy ranges.
} \label{fig:current}
\end{figure}
 
In Figure~\ref{fig:current}, we plot the currents $ \langle \hat{j}_{\uparrow, \downarrow} \rangle_n$ as a function of the eigenenergies $E_n$, as calculated for the isotropic two-leg ladder in panel (a) and for the anisotropic ladder [Eq.~\eqref{effective_ham_2D}] in panel (b); blue and red circles correspond to $ \langle \hat{j}_{\uparrow} \rangle$ and $ \langle \hat{j}_{\downarrow} \rangle$, respectively. We consider a region of 201 lattice sites along each ladder, corresponding in the anisotropic model to the region $\lambda\! \in\! \{0,..., 200\}$. Note that the isotropic results can be recovered from the anisotropic ladder [Eq.~\eqref{effective_ham_2D}] if the finite region $\mathcal{R}$ is
translated to sufficiently higher $\lambda$. 
All quantities are expressed in terms of the energy scale $J_y$, which is set to $J_y\!=\!J/2$ in the isotropic limit, and $J_y \!=\! J_{\lambda\!=\!200}/2$ in the anisotropic case. In both panels of Fig.~\ref{fig:current}, there are large energy regimes over which the current smoothly varies as a function of energy. These are the states which will turn into the usual quantum Hall edge modes upon adding more legs~\cite{Celi:2014}. Most importantly, chiral currents persist in this model even in the presence of anisotropy. Finally, in the centre and outer edges of the energy spectrum $E_n$, there are regions where the variation of the current is less smooth; this corresponds to the regions where the bulk states can be found in the 2D limit~\cite{Celi:2014}.

\end{document}